\documentclass[12pt,a4paper, reqno]{amsart}
\usepackage[utf8]{inputenc}
\usepackage[english]{babel}
\usepackage{amsmath,mathrsfs,amsfonts,amssymb,amsxtra,latexsym,amscd,amsthm,marvosym,multirow,mathtools, color, stmaryrd}
\usepackage[a4paper,inner=2.3cm, outer=2.3cm, top=2.3cm, bottom=2.3cm]{geometry}
\usepackage{eucal}
\usepackage{enumerate}
\usepackage[shortlabels]{enumitem}
\usepackage{csquotes}
\usepackage{hyperref}

\newtheorem{theorem}{Theorem}[section]
\newtheorem{lemma}[theorem]{Lemma}
\newtheorem{assumption}[theorem]{Assumption}
\newtheorem{proposition}[theorem]{Proposition}

\theoremstyle{remark}
\newtheorem{notation}[theorem]{Notation}
\newtheorem{remark}[theorem]{Remark}
\newtheorem{definition}[theorem]{Definition}
\newtheorem{example}{Example}

\renewcommand{\Re}{\mathrm{Re}\,}
\renewcommand{\Im}{\mathrm{Im}\,}

\newcommand{\N}{\mathbb{N}}
\newcommand{\Z}{\mathbb{Z}}
\newcommand{\R}{\mathbb{R}}
\newcommand{\C}{\mathbb{C}}
\newcommand{\A}{\textbf{A}}
\newcommand{\B}{\textbf{B}}
\newcommand{\dd}{\mathrm{d}}

\makeatletter

\@addtoreset{equation}{section}  
\makeatother

\usepackage[normalem]{ulem}
\definecolor{DarkGreen}{rgb}{0,0.5,0.1} 

\newcommand\soutD{\bgroup\markoverwith
{\textcolor{DarkGreen}{\rule[.5ex]{2pt}{1pt}}}\ULon}
\newcommand{\Hm}[1]{\leavevmode{\marginpar{\tiny%
$\hbox to 0mm{\hspace*{-0.5mm}$\leftarrow$\hss}%
\vcenter{\vrule depth 0.1mm height 0.1mm width \the\marginparwidth}%
\hbox to
0mm{\hss$\rightarrow$\hspace*{-0.5mm}}$\\\relax\raggedright #1}}}

\renewcommand{\leq}{\leqslant}	\renewcommand{\geq}{\geqslant}

\begin{document}
\title[]{The Laplacian with complex magnetic fields}
\dedicatory{Dedicated to our colleague and friend Jussi Behrndt
on the occasion of his 50th birthday}
\author[D. Krej\v{c}i\v{r}\'{i}k]{David Krej\v{c}i\v{r}\'{i}k}
\address[David Krej\v{c}i\v{r}\'{i}k]{Department of Mathematics, Faculty of Nuclear Sciences and Physical Engineering, Czech Technical University in Prague, Trojanova 13, 12000 Prague, Czechia.}
\email{david.krejcirik@fjfi.cvut.cz}
\author[T. Nguyen Duc]{Tho Nguyen Duc}
\address[Tho Nguyen Duc]{Analytical and Algebraic Methods in Optimization Research Group, Faculty of Mathematics and Statistics, Ton Duc Thang University, Ho Chi Minh City, Vietnam.}
\email{nguyenductho@tdtu.edu.vn}

\author[N. Raymond]{Nicolas Raymond}
\address[Nicolas Raymond]{Univ Angers, CNRS, LAREMA, Institut Universitaire de France, F-49000 Angers, France}
\email{nicolas.raymond@univ-angers.fr}
\begin{abstract}
We study the two-dimensional magnetic Laplacian when the magnetic field is allowed to be complex-valued. Under the assumption that the imaginary part of the magnetic potential is relatively form-bounded with respect to the real part of the magnetic Laplacian, we introduce the operator as an m-sectorial operator. In two dimensions, sufficient conditions are established to guarantee that the resolvent is compact. In the case of non-critical complex magnetic fields, a WKB approach is used to
construct semiclassical pseudomodes, 
which do not exist when the magnetic field is real-valued.  
\end{abstract}
\maketitle

\section{Introduction}

Since the appearance of the highly influential paper
\cite{Abramov-Aslanyan-Davies_2001} at the turn of the millennium,
there has been a growing interest in Schr\"odinger operators 
with complex-valued electric potentials.
This is explained not only by diverse physical motivations, 
but notably due to a new concept of quantum mechanics 
where observables can be represented by non-self-adjoint operators 
\cite{GHS,KSTV15}. 
Moreover, the mathematical studies lead to 
unprecedented spectral properties such as 
the existence of pseudomodes 
\cite{Davies_1999-NSA,DSZ04,KS19,Arnal-Siegl_2023}
or the existence of eigenvalues accumulating at non-zero points
of the essential spectrum 
\cite{Bogli_2017,Cuenin_2022,Bogli-Cuenin_2023}.

It is about time that someone would address 
the case of Schr\"odinger operators
where the magnetic potential is complex-valued now.
This is the subject of the present paper. We are not merely motivated by pure mathematical curiosity, but also by important physical considerations. 
Among these, let us mention 
superconductors \cite{Hatano-Nelson_1996,Hatano-Nelson_1998},
quantum statistical physics 
\cite{Peng-etal_2015, Ananikian-Kenna_2015},
speculations about a novel type of magnetic monopoles~\cite{James-24},
stability of black holes in general relativity 
\cite{Jaramillo_2015a,Jaramillo_2015b}
and the concept of quasi-self-adjointness again~\cite{Krejcirik-19}.
What is more, 
making the magnetic field complex is mathematically 
much more challenging than its electric counterpart.
That is probably why rigorous results do not exist in the literature
and the objective of this paper is to fill in this gap.

\subsection{The problem}
We are concerned with the differential expression
\begin{equation}\label{Magnetic Sch}
	\left(-ih\nabla -\A \right)^2
	=\sum_{j=1}^d\left(-ih\partial_{x_{j}}-A_{j}\right)^2
	\,,
\end{equation}
where $d \geq 1$ is the dimension,
$\A= (A_{1},\ldots, A_d):\R^d \to \C^d$ 
is the complex (magnetic) vector potential 
and~$h$ is a small positive (semiclassical) parameter.
If the imaginary part of~$\A$ is non-zero,
it is not even clear that~\eqref{Magnetic Sch}
leads to a well defined operator in $L^2(\R^d)$.
More specifically, our first concern is
\begin{enumerate}
\item[(I)]
to identify the right subspace of $L^2(\R^d)$ as a domain
on which~\eqref{Magnetic Sch} is realised 
as a closed operator with non-empty resolvent set.
\end{enumerate}
Our next curiosity is about
complex ``magnetic bottles'',
\emph{i.e.},
\begin{enumerate}
\item[(II)]
to find conditions on~$\A$
which guarantee that the operator has compact resolvent.
\end{enumerate}
Since it is the pseudospectrum which describes 
non-self-adjoint phenomena, our last task is 
\begin{enumerate}
\item[(III)]
to construct pseudomodes in the semiclassical limit $h \to 0$. 
\end{enumerate}
Of course, other problems could be raised,
but already these three tasks lead 
to considerable mathematical challenges.

\subsection{Main results}
\subsubsection{The magnetic Laplacian as an $m$-sectorial operator}
In this paper, we deal with task~(I) by assuming 
that the imaginary part of the vector potential 
is relatively form-bounded with respect to 
the real part of the magnetic Laplacian.
More specifically, we always make the following hypothesis.
\begin{assumption}\label{assumption.1}
	Let $\A\in L^{2}_{\mathrm{loc}}(\R^d, \C^d)$ 
	and assume that there exist two constants $a\in (0,1)$ and $b\geq 0$ such that
	\begin{equation}\label{Eq Sectorial cond}
		\int_{\R^d} \vert (\Im \A) u \vert^2 \, \dd x\leq a\int_{\R^d} \vert(-ih\nabla -\Re \A)u \vert^2\, \dd x +b \int_{\R^d} \vert u \vert^2\, \dd x\,,
	\end{equation}
	for all $u\in C_{c}^{\infty}(\R^d)$.
\end{assumption}
We define the sesquilinear form naturally associated with~\eqref{Magnetic Sch}, \emph{i.e.},
\begin{align}
		Q_{h,\A}(u,v) 
		&:= \int_{\R^d} 
		(-ih\nabla -\A)u\cdot \overline{(-ih\nabla -\overline{\A})v} \ \dd x
		\,,
		\label{form}
		\\
		\mathrm{Dom}(Q_{h,\A})
		&:= \left\{ u\in L^2(\R^d):
		(-ih\nabla-\Re \A)u \in L^2(\R^d,\C^d),
		\, (\Im \A)u \in L^2(\R^d,\C^d)\right\}
		\,.
		\nonumber
\end{align}
The objective of task~(I) is fulfilled by the following theorem.
\begin{theorem}\label{Thrm Operator}
Under Assumption~\ref{assumption.1}, $Q_{h,\A}$ is densely defined, closed and sectorial. Then, the operator defined by
\begin{equation*}
		\begin{aligned}
		\mathrm{Dom}(\mathscr{L}_{h,\A}) \coloneqq & \left\{
			u \in \mathrm{Dom}(Q_{h,\A}) \hspace{0.2 cm}:\hspace{0.2 cm} \exists f\in L^2(\R^d),\hspace{0.2 cm} \forall v\in \mathrm{Dom}(Q_{h,\A}), \hspace{0.2 cm} Q_{h,\A}(u,v) = \langle f,v \rangle \right\}\,,\\
			\mathscr{L}_{h,\A}u \coloneqq & f\,,
		\end{aligned}
	\end{equation*}
is m-sectorial.
\end{theorem}

While the closed representative $\mathscr{L}_{h,\A}$ of~\eqref{Magnetic Sch} 
is easily introduced in all dimensions, 
it is well known that its spectral analysis
becomes cumbersome in higher dimensions even in the self-adjoint case.
Therefore, we modestly restrict to dimension $d=2$ in the sequel.
In this case, the magnetic field
\[
  \B
  \coloneqq \operatorname{curl} \A
  = \partial_{x_{1}}A_{2}-\partial_{x_{2}}A_{1}
\]
is a scalar function.
Then pointwise sufficient conditions which guarantee 
Assumption \ref{assumption.1} are contained 
in the following proposition.

\begin{proposition}\label{Prop Ass on B}
	Let $\A \in L^2_{\textup{loc}}(\R^2,\C^2)$ 
	and $\B\in L^1_{\textup{loc}}(\R^2)$. 
	Assume one of the following conditions:
	\begin{enumerate}[label=\textup{\textbf{(C\arabic*):}}, ref=\textup{\textbf{C\arabic*}}]
		\item \label{bound by Re B} there exist $\varepsilon_{1} \in (0,1)$, $C_{1}\in \R$ such that
		\begin{align*}
			|\Im \A (x)|^2\leq  \pm \varepsilon_{1}h\Re \B(x) +C_{1}\,,\qquad \forall x\in \R^2\,;
		\end{align*}
		\item \label{bound by Im B}  there exist $\varepsilon_{2} \in \left(0,\frac{1}{2}\right)$, $C_{2}\in \R$ such that
		\[ |\Im \A (x)|^2\leq  \pm \varepsilon_{2}h\Im \B(x) +C_{2}\,,\qquad \forall x\in \R^2\,.\]
	\end{enumerate}
	Then Assumption \ref{assumption.1} holds.
\end{proposition}

Of course, the conditions are automatically satisfied if $\Im\A = 0$ and if $\Re\mathbf{B}$ is bounded from below or from above. If $ \Im\mathbf{B}=0$, note that one can choose a real vector potential $\mathbf{A}$, but that one can also choose a complex gauge (leading to non unitarily equivalent operators).

In practice, when $\A$ and $\B$ are continuous on $\R^2$, it suffices to verify \eqref{bound by Re B} or \eqref{bound by Im B} for large values of $|x|$. Furthermore, if $\Im \A$ is bounded and $\Re \B(x)$ (respectively, $\Im \B(x)$) does not change sign for large $|x|$, then \eqref{bound by Re B} (respectively, \eqref{bound by Im B}) holds.  

\begin{example}[Homogeneous magnetic field] 
Unfortunately, Assumption~\ref{assumption.1}
excludes the important case of constant magnetic field
$\B = c \in\C$ corresponding to 
\[ 
  \A(x) := 
  c \, \mbox{$\frac{1}{2}$} \left( -x_2, x_1\right)
  \qquad \mbox{or} \qquad
  \A(x) := 
  c \left( 0, x_1\right)
\,  ,
\]
unless $\Im c = 0$.
In fact, implementing task~(I) in this case seems particularly non-trivial.
\end{example}
\begin{example}[Complex Miller--Simon's potential] 
Consider
\[ 
  \A(x) := 
  \left( -\frac{cx_{2}}{(1+|x|)^{\alpha}}, \frac{cx_{1}}{(1+|x|)^{\alpha}}\right)
  \,,
\]
where $c \coloneqq c_{1}+ic_{2}\in \C$ with $c_1,c_2 \in \R$ and $\alpha>0$. 
When $c$ is real, the magnetic Laplacian with this type of magnetic potential was considered in~\cite{Miller-Simon-80}. When $c_{2}\neq 0$ and if we assume further that $\alpha\geq 1$ then it can be checked that both condition  \eqref{bound by Re B} and \eqref{bound by Im B} are satisfied. Indeed, since $|\A|$ is bounded and that
\[ 
  \B(x) 
  =c \left(\frac{2}{(1+|x|)^\alpha} - \frac{\alpha |x|}{(1+|x|)^{\alpha+1}} \right)
  \]
is also bounded on $\R^2$, we can therefore find sufficient large constants $C_{1}$ and $C_{2}$ such that  \eqref{bound by Re B} and \eqref{bound by Im B} hold. With more effort, we can see that $\mathrm{Dom}(Q_{h,\A})=H^1(\R^2)$ and $\operatorname{Dom}(\mathscr{L}_{h,\A})=H^2(\R^2)$. In particular, when $c_{1}=0$, we have an example of purely imaginary magnetic fields such that our magnetic Laplacian is well defined.
\end{example}
\begin{example}{(Purely imaginary exponential magnetic potentials)}\label{Ex.exponential}
Now we give an example of an unbounded purely imaginary $\A$ 
for which \eqref{bound by Re B} fails but \eqref{bound by Im B} still holds. 
Consider
\begin{equation}\label{Exponential}
\A(x) \coloneqq i c e^{|x|^2} (-x_{2},x_{1})\,, 
\end{equation}
where $c \in (-h,h)$ and $h>0$.
It can be checked that
\[ |\Im \A(x)|^2 = c^2 |x|^2 e^{|x|^2},\qquad 
\Im \B(x) = 2c(|x|^2+1)e^{|x|^2}\,.\]
Hence, we have
\[ |\Im \A(x)|^2 \leq \frac{c}{2} \Im \B(x),
\qquad \forall x\in \R^2\,. \]
By choosing $C_{2}=0$ and $\varepsilon_{2} \in \left(\frac{|c|}{2h},\frac{1}{2}\right)$, \eqref{bound by Im B} is verified.
\end{example}

\subsubsection{Compactness of the resolvent}
Let us now consider task~(II) about the compactness of the resolvent 
(which implies that the spectrum is purely discrete).

\begin{theorem}\label{Thrm Copact Res}
	Let $\A$ satisfy Assumption~\ref{assumption.1} 
	and $\Re \A \in L^{\infty}_{\textup{loc}}(\R^2,\R^2)$. 
	Assume one of the following conditions:
	\begin{enumerate}[label=\textup{\textbf{(H\arabic*):}}, ref=\textup{\textbf{H\arabic*}}]
		\item \label{Re B to inf} $\Re \B\in C^{0}(\R^2)$  and $\displaystyle \lim_{\vert x \vert \to+\infty} \vert \Re \B(x) \vert = +\infty$\,,
		\item \label{Im B to inf} $\Im \B\in C^{0}(\R^2)$  and $\displaystyle \lim_{\vert x \vert \to+\infty} \vert \Im \B(x) \vert = +\infty$\,,
		\item \label{Im A to inf} $\Im \A \in C^{0}(\R^2,\R^2)$  and $\displaystyle\lim_{\vert x \vert \to+\infty} \vert \Im \A(x) \vert = +\infty$\,.
	\end{enumerate}
	Then $\mathscr{L}_{h,\A}$ has compact resolvent.
\end{theorem}

The first condition of the theorem extends 
the usual magnetic bottle realisation \cite{Iwatsuka_1986}
to the non-self-adjoint setting.
More interestingly, it follows that the magnetic Laplacian of Example~\ref{Ex.exponential} (with $c\neq 0$)
has compact resolvent despite $\Re \B=0$.

\begin{example}
We can construct examples when $\mathbf{(H2)}$ holds but $\mathbf{(H3)}$ does not. Consider
	\[ A_{1}(x)=0, \qquad A_{2}(x) = x_{1}\left(\frac{x_{1}^8}{9}+x_{2}^{8} \right)+ i x_{1}\left(\frac{x_{1}^2}{3}+x_{2}^{2} \right).\]
	Then, the corresponding magnetic field is 
	\[ \mathbf{B}(x) = (x_{1}^{8}+x_{2}^{8})+i(x_{1}^2+x_{2}^2).\]
	We compute
	\[ \vert \textup{Im} \mathbf{A}(x) \vert^2=x_{1}^2\left(\frac{x_{1}^2}{3}+x_{2}^{2} \right)^2.\]
	It can be verified that condition $\mathbf{(C1)}$ is satisfied, and hence Assumption 1.1 holds. Clearly, $\mathbf{(H2)}$ is satisfied, but $\mathbf{(H3)}$ is not since $\vert \textup{Im} \mathbf{A}(0,x_2) \vert =0$.
	\end{example}

\subsubsection{Semiclassical pseudomodes}
Finally, let us present our implementation of task~(III)
about the construction of semiclassical pseudomodes.
To this purpose, let us restrict to
magnetic potentials $\A\in C^{\infty}(\R^2,\C^2)$ 
satisfying Assumption~\ref{assumption.1}.
In the plane~$\R^2$, we define the subset 
\begin{equation*}
	\Gamma := \left\{ x\in \R^2 \left| \
	\begin{aligned}
	&\Im \A (x)=0, \hspace{0.2 cm} \B(x)\neq 0, \hspace{0.2 cm} \partial_{\overline{z}} \B(x)\neq 0,\\
	 &Q_{1}(x)>0, \hspace{0.2 cm} Q_{1}(x)Q_{3}(x)-Q_{2}^2(x)>0
	\end{aligned}
		 \right.
	\right\}\,,
\end{equation*}
where
\begin{equation}\label{Q123}
\begin{aligned}
			&Q_1(x) := \frac{1}{4} \textup{Re}\,\left[ \B(x)\left(1+\frac{\partial_{z} \B}{\partial_{\overline{z}} \B}(x)\right)\right]+\frac{1}{2}\partial_{x_{1}}\Im A_{1}(x)\,,\\
			&Q_2(x) := \frac{1}{4} \mathrm{Im}\, \left[ \B(x)\frac{\partial_{z} \B}{\partial_{\overline{z}} \B}(x)\right]+\frac14\left(\partial_{x_{1}} \Im A_{2}+\partial_{x_{2}}\Im A_{1}(x)\right)\,,\\
			&Q_3(x) :=  \frac{1}{4} \textup{Re}\,\left[ \B(x)\left(1-\frac{\partial_{z} \B}{\partial_{\overline{z}} \B}(x)\right)\right]+\frac{1}{2}\partial_{x_{2}}\Im A_{2}(x)\,,
		\end{aligned}
\end{equation}
and 
$
\partial_{z}:= \frac{1}{2} 
\left( \partial_{x_1}-i\partial_{x_2}\right)
$
and
$
\partial_{\overline{z}}:= \frac{1}{2} 
\left( \partial_{x_1}+i\partial_{x_2}\right)
$
stand for the usual Wirtinger derivatives. Here, $Q_{1}$, $Q_{2}$ and $Q_{3}$ are determined by both the magnetic field and the imaginary part of the magnetic potential up to their first derivatives. At a point on $\Gamma$, these functions act as coefficients in a positive definite quadratic form that governs how the pseudomode decays as we move away from that point.
Our main result is the following theorem.

\begin{theorem}\label{Theorem exponential}
	Let $x^{0}\in \Gamma$ and assume that $\B$ is real-analytic at $x^{0}$. 
	Then there exist constants $C,h_0>0$ and a family of functions 
	$\left( u_h \right)_{0<h\leq h_{0}}\subset C_{c}^{\infty}(\R^2)$ 
	such that, for all $h\in(0, h_{0})$,
	\begin{equation}\label{Exp Decay}
		\left\Vert \left( \mathscr{L}_{h,\A} - h\B(x^{0})\right)u_h\right\Vert \leq  \exp\left(-\frac{C}{h^{1/7}}\right) \Vert u_h \Vert \,.
	\end{equation}
\end{theorem}

Note that $\Gamma = \varnothing$ if $\Im\A = 0$,
because $Q_{1}Q_{3}-Q_{2}^2 = 0$ in this case.
It follows that the existence of the pseudomode 
$\left( u_h \right)_{0<h\leq h_{0}}$ is possible 
only in the present non-self-adjoint setting. 
Below we provide examples of magnetic potentials 
for which $\Gamma \not= \varnothing$;
these include polynomial and oscillating complex magnetic fields.

In order to prove Theorem~\ref{Theorem exponential}, 
we use the ideas of the magnetic WKB strategy as developed 
in \cite{BR-20,GNRV-21} in the self-adjoint case 
to find approximations of eigenfunctions.
In our non-self-adjoint development, it is remarkable that
the magnetic potential~$\A$ does not need to be analytic 
for this result to hold. This technical outcome is due to the use of 
a formal gauge, which allows for the local transformation 
to an appropriate analytic potential at $x^{0}$, see Section~\ref{Choice MP}. The power $h^{\frac{1}{7}}$ in \eqref{Exp Decay} arises from the transport solutions estimate in Lemma \ref{lem.boundaj}, which intrinsically reflects the order of the partial derivatives involved in the recursive formula \eqref{Transport Solution j+1}. However, inspired by the exponential decay $\mathcal{O}(e^{-C/h})$ in the Schr\"{o}dinger case \cite[Theo. 1.1]{DSZ04}, it is natural to ask whether the power of $h$ in \eqref{Exp Decay} is optimal.

We stress that Theorem~\ref{Theorem exponential} 
is not covered by~\cite{DSZ04} (see also \cite{Z01} and the seminal work \cite[Chapter 27]{Hormander}).
Indeed, the Weyl symbol of $\mathscr{L}_{h,\A}$ is
\[
p(x,\xi) =|\xi-\Re\mathbf{A}|^2-|\Im\mathbf{A}|^2-2i\langle\xi-\Re\mathbf{A},\Im\mathbf{A}\rangle\,.
\] 
Note that $p(x,\xi)=0$ is equivalent to 
$\xi-\Re\mathbf{A}=\pm(\Im\mathbf{A})^{\perp}$
and that
\[\begin{split}
\{&\Re p,\Im p \}(x,\xi)\\ &=-4(\xi-\Re\mathbf{A})\cdot\partial_{x}\langle\xi-\Re\mathbf{A},\Im\mathbf{A}\rangle-2\partial_{x}(|\xi-\Re\mathbf{A}|^2-|\Im\mathbf{A}|^2)\cdot\Im\mathbf{A}\,,
\end{split}
\]
where $\{ \cdot, \cdot\}$ is the Poisson bracket. 
Given $x^{0}\in \Gamma$, then for any point $(x^{0},\xi^{0})\in \R^{4}$ such that $p(x^{0},\xi^{0})=0$, we have $\{\Re p,\Im p \}(x^{0},\xi^{0})=0$ (since $\Im\mathbf{A}(x^0)=0$), so \cite[Thm.~1.2]{DSZ04} does not apply. However, there might be hope to weaken the Poisson bracket condition as in \cite{PS04} or in \cite[Example 5.4]{KS19} where the electric Schrödinger operator is considered.

\subsection{Structure of the paper}
In Section~\ref{Sec Definition} we deal with tasks~(I) and~(II);
in particular, we introduce the operator~$\mathscr{L}_{h,\A}$
and prove Proposition~\ref{Prop Ass on B} and Theorem~\ref{Thrm Copact Res}.
Task~(III) is considered in Section~\ref{Sec WKB}; 
namely, we establish Theorem~\ref{Theorem exponential}
and provide the specific examples which the theorem applies to.
%
 
\section{Sectorial magnetic Laplacians}\label{Sec Definition}
%
The main purpose of this section is to realise 
the differential expression~\eqref{Magnetic Sch}
as an m-sectorial operator in $L^2(\R^d)$ via a sesquilinear form and study its  compact resolvent property.

Assume $\A\in L^2_{\textup{loc}}(\R^d,\C^d)$. Expanding $(-ih\nabla-\mathbf{A})^2$, the action of this operator should be  
\begin{align*}
(-ih\nabla-\Re\mathbf{A})^2-(\Im \mathbf{A})^2-i\left(\Im \mathbf{A} \cdot(-ih\nabla-\Re\mathbf{A})+(-ih\nabla-\Re\mathbf{A})\cdot\Im\mathbf{A}\right)\,.
\end{align*}
Aiming at the variational definition, 
this suggests considering the form domain
\begin{equation}\label{Form domain}
	\mathcal{V}_{h,\A}:= \left\{ u\in L^2(\R^d):(-ih\nabla-\Re \A)u \in L^2(\R^d,\C^d),\, (\Im \A)u \in L^2(\R^d,\C^d)\right\}
\end{equation}
equipped with the natural inner product
\begin{equation}\label{Inner product}
	\langle u, v \rangle_{\mathcal{V}_{h,\A}}:= \langle u, v\rangle+\langle (-ih\nabla-\Re \A)u, (-ih\nabla-\Re \A)v \rangle+\langle (\Im \A)u, (\Im \A)v \rangle\,,
\end{equation}
where $\langle\cdot,\cdot\rangle$ denotes the inner product of $L^2(\R^d)$,
linear in the first component.

\begin{remark}
Since $\A\in L^2_{\textup{loc}}(\R^d,\C^d)$, it follows that for all $u\in L^2(\R^d)$,
\[(\Re \A) u\in L^1_{\textup{loc}}(\R^d,\C^d)\,,\quad (\Im \A) u\in L^1_{\textup{loc}}(\R^d,\C^d)\,.\]  
In particular, $(-ih\nabla -\Re \A)u$ and $(\Im \A) u$ can be understood in the sense of distributions in \eqref{Form domain}.
\end{remark}
Just as in the self-adjoint case, we have the following properties of the form domain.
\begin{proposition}\label{Prop V}
Let $h>0$ and $\A\in L^2_{\textup{loc}}(\R^d,\C^d)$, the following holds.
\begin{enumerate}[label=\textbf{\textup{(\alph*)}}]
\item \label{Hilbert}$\left(\mathcal{V}_{h,\A}, \langle \cdot, \cdot \rangle_{\mathcal{V}_{h,\A}}\right)$ is a Hilbert space.
\item \label{Dense}$C_{c}^{\infty}(\R^d)$ is dense in $\mathcal{V}_{h,\A}$.
\item \label{H1} Under the assumption \ref{assumption.1}, $ \mathcal{V}_{h,\A} =  \mathcal{V}_{h,\Re \A}$, and two norms $\Vert \cdot \Vert_{\mathcal{V}_{h,\Re \A}}$ and $\Vert \cdot \Vert_{\mathcal{V}_{h,\A}}$ are equivalent.
\end{enumerate}
\end{proposition}

\begin{proof}
The proof of~\ref{Hilbert} is standard, and the proof of the density result in~\ref{Dense} follows the same steps as in \cite[Theorem 7.22]{LL01}, so we omit the details here. Using~\ref{Dense}, we extend the assumption \eqref{Eq Sectorial cond} to hold on the space $\mathcal{V}_{h,\A}$. Consequently, statements in \ref{H1} are established.
\end{proof}
After exploring several key properties of the space $\mathcal{V}_{h,\A}$, we are now ready to prove our first main theorem.
\begin{proof}[Proof of Theorem \ref{Thrm Operator}]
We recall the sesquilinear form $Q_{h,\A}: \mathcal{V}_{h,\A} \times \mathcal{V}_{h,\A} \to \C$ defined in ~\eqref{form}. It is evident that $Q_{h,\A}$ is densely defined. Writing $\A=\Re \A+i \Im \A$, we expand $Q_{h,\A}$ as follows
\begin{align*}
Q_{h,\A}(u,u) 
		= &\int_{\R^d} (-ih\nabla -\Re \A-i\Im \A)u\cdot 
		\overline{(-ih\nabla -\Re \A+i\Im \A)u}\, \dd x\\
		= &\int_{\R^d}\left(|(-i h \nabla-\Re \A)u|^2-|\Im \A|^2 |u|^2\right)\dd x-2i\Re\langle\Im \A\, u, (-i h \nabla-\Re \A)u\rangle\,.
\end{align*}
Under the assumption \eqref{Eq Sectorial cond} (which extends to $\mathcal{V}_{h,\A}$ by the density of $C_{c}^{\infty}(\R^{d})$ in $\mathcal{V}_{h,\A}$), we obtain that for all $u\in \mathcal{V}_{h,\A}$,
\begin{equation}\label{eq.ReQ>}
\Re Q_{h,\A}(u,u)\geq (1-a)\int_{\R^d}|(-i h \nabla-\Re \A)u|^2\dd x-b\|u\|^2\,,
\end{equation}
where $a \in (0, 1)$ and $b \geq 0$ are constants from \eqref{Eq Sectorial cond}.

Next, we estimate the imaginary part of $Q_{h,\A}(u,u)$:
\begin{align*}
|\Im Q_{h,\A}(u,u)|&\leq2\|\Im \A\,u\|\|(-i h \nabla-\Re \A)u\|\\
&\leq \|\Im\mathbf{A}u\|^2+\|(-ih\nabla-\mathbf{A})u\|^2\\
&\leq (1+a)\|(-ih\nabla-\mathbf{A})u\|^2+b\|u\|^2\,.
\end{align*}
Combining these inequalities, we deduce
\[|\Im Q_{h,\A}(u,u)|\leq \frac{1+a}{1-a} \Re Q_{h,\A}(u,u) + \frac{2b}{1-a} \Vert u \Vert^2 \,.\]
This shows that the form $Q_{h,\A}$ is sectorial. 

We now verify the closedness of the form. Consider the form 
$$t(u,v) \coloneqq \langle (-ih\nabla -\Re \A)u , (-ih\nabla -\Re \A) v \rangle\,,$$ which is sectorial and closed on $ \mathcal{V}_{h,\Re \A}$. By applying \cite[Thm. VI.1.33]{Kato95} and noting that the form $\langle \Im \A \cdot, \Im \A \cdot \rangle$ is $t$-bounded, we conclude that the form $\Re Q_{h,\A}$ is closed. By a remark in \cite[Sec. VI.1.3]{Kato95}, the sectorial form $Q_{h,\A}$ is also closed.

Consequently, $Q_{h,\A}$~gives rise to
an m-sectorial operator $\mathscr{L}_{h,\A}$
via the standard representation theorem \cite[Thm.~VI.2.1]{Kato95}.
\end{proof}

From now on, we restrict to dimension $d=2$. In the following lemma, we establish two magnetic inequalities corresponding to the real and imaginary parts of $\B$. The first inequality is well known, see \cite[Theo. 2.9]{AHS-78}.
\begin{lemma}\label{Lem Mag Ineqs}
Let $\A \in L^2_{\textup{loc}}(\R^2,\C^2)$
and $\B\in L^1_{\textup{loc}}(\R^2)$. Then, for all $u\in C_{c}^{\infty}(\R^2)$, we have
	\begin{align*}
		\left\vert \int_{\R^2} h\Re \B |u|^2\, \dd x \right\vert 
		&\leq \int_{\R^2} \left\vert (-ih\nabla-\Re \A) u \right\vert^2\, \dd x\,,
		\\
		\left\vert \int_{\R^2} h\Im \B |u|^2\, \dd x \right\vert 
		&\leq \int_{\R^2} \left\vert (-ih\nabla-\Re \A) u \right\vert^2\, \dd x+\int_{\R^2} \left\vert (\Im \A) u \right\vert^2\, \dd x\,.
	\end{align*}
\end{lemma}
\begin{proof}
			Since $\A \in L^{2}_{\textup{loc}}(\R^2,\C^2)$ and $\B\in L^{1}_{\textup{loc}}(\R^2)$, all the integrals appearing in the statements of this lemma are well defined. Let $u\in C_{c}^{\infty}(\R^2)$. By observing that
			\[ \left[(-ih\partial_{x_{1}}-A_{1}),\,(-ih\partial_{x_{2}}-A_{2})\right]=ih\B\,,\]
			we have
			\[ ih\B |u|^2 = \overline{u}(-ih\partial_{x_{1}}-A_{1})(-ih\partial_{x_{2}}-A_{2})u-\overline{u}(-ih\partial_{x_{2}}-A_{2})(-ih\partial_{x_{1}}-A_{1})u\,.\]
			Integrating by parts, we get
			\begin{align*}
				\int_{\R^2} ih\B |u|^2\, \dd x = \langle (-ih\partial_{x_{2}}-A_{2})u, (-ih\partial_{x_{1}}-\overline{A_{1}})u\rangle -\langle (-ih\partial_{x_{1}}-A_{1})u, (-ih\partial_{x_{2}}-\overline{A_{2}})u\rangle \,.
			\end{align*}
			By writing out the real and imaginary parts of $A_{1}$ and $A_{2}$, we find that
			\begin{align*}
				\int_{\R^2} ih\B |u|^2\, \dd x=& \ 2\Im \langle (\Im A_{2})u,(-ih\partial_{x_{1}}-\Re A_{1})u\rangle -2\Im \langle (\Im A_{1})u,(-ih\partial_{x_{2}}-\Re A_{2})u\rangle\\
				&+i 2\Im \langle (-ih\partial_{x_{2}}-\Re A_{2})u, (-ih\partial_{x_{1}}-\Re A_{1})u\rangle\,.
			\end{align*}
			In other words, we have
			\[\int_{\R^2} h\Re\B |u|^2\, \dd x = 2\Im \langle (-ih\partial_{x_{2}}-\Re A_{2})u, (-ih\partial_{x_{1}}-\Re A_{1})u\rangle\,, \]
			and
			\begin{align*}
				\int_{\R^2} h\Im \B |u|^2\, \dd x = &2\Im \langle (\Im A_{1})u,(-ih\partial_{x_{2}}-\Re A_{2})u\rangle\\
				&-2\Im \langle (\Im A_{2})u,(-ih\partial_{x_{1}}-\Re A_{1})u\rangle\,.
			\end{align*}
			The conclusion follows from the Cauchy--Schwarz inequality.
\end{proof}

Now we are in a position to establish Proposition~\ref{Prop Ass on B}.
\begin{proof}[Proof of Proposition~\ref{Prop Ass on B}]
When $\Im \A$ and $\Re \B$ satisfy \eqref{bound by Re B}, then Assumption \ref{assumption.1} is a direct consequence of the first inequality in Lemma \ref{Lem Mag Ineqs}. 
	
	Now, we assume that $\Im \A$ and $\Im \B$ satisfy \eqref{bound by Im B} with the plus sign (for instance). From the second inequality in Lemma \ref{Lem Mag Ineqs},
	\begin{align*}
		\int_{\R^2} \vert (\Im \A) u \vert^2 \, \dd x &\leq \varepsilon_{2} h\int_{\R^2} \Im \B |u|^2\, \dd x+C_{2} \Vert u \Vert^2\\
		&\leq  \varepsilon_{2}\int_{\R^2} \left\vert (-ih\nabla-\Re \A) u \right\vert^2\, \dd x+\varepsilon_{2}\int_{\R^2} \left\vert (\Im \A) u \right\vert^2\, \dd x+C_{2} \Vert u \Vert^2\,.
	\end{align*}
	This implies that Assumption \ref{assumption.1} holds with $a=\frac{\varepsilon_2}{1-\varepsilon_2}\in(0,1)$ and $b=\frac{C_2}{1-\varepsilon_2}$.
\end{proof}

Now we turn to task~(II) about the compactness 
of the resolvent of~$\mathscr{L}_{h,\A}$.
\begin{proof}[Proof of Theorem \ref{Thrm Copact Res}]
Thanks to \cite[Prop.~4.24]{CR21}, it is equivalent to proving that the injection 
\[ \left(\mathrm{Dom}(\mathscr{L}_{h,\A}), \Vert \cdot \Vert_{\mathscr{L}_{h,\A}}\right) \hookrightarrow \left( L^2(\R^2), \Vert \cdot \Vert\right)\]
is compact, where $\Vert \cdot \Vert_{\mathscr{L}_{h,\A}}\coloneqq\Vert \mathscr{L}_{h,\A} \cdot \Vert + \Vert \cdot \Vert$ is the graph norm. From \eqref{eq.ReQ>} and Proposition \ref{Prop V}\ref{H1}, there exist $\gamma>0$ and $\mu>0$  such that for all $u\in\mathrm {Dom}(\mathscr{L}_{\mathbf{A}})$, we have
\[\gamma \Vert u \Vert_{\mathcal{V}_{h,\A}}^2\leq \vert Q_{h,\A}(u,u)\vert + \mu \Vert u \Vert^2\leq \Vert (\mathscr{L}_{h,\A}+\mu) u \Vert \Vert u \Vert \leq \left(\frac{1}{2}+\mu\right)\Vert u \Vert_{\mathscr{L}_{h,\A}}^2 \,,\]
which shows that the injection
\[\left(\mathrm{Dom}(\mathscr{L}_{h,\A}), \Vert \cdot \Vert_{\mathscr{L}_{h,\A}}\right)\hookrightarrow \left(\mathcal{V}_{h,\A}, \Vert \cdot \Vert_{\mathcal{V}_{h,\A}}\right)\]
is continuous. Since the space of compact operators forms an ideal within the space of bounded operator, it remains to explain why $\left(\mathcal{V}_{h,\A}, \Vert \cdot \Vert_{\mathcal{V}_{h,\A}}\right)$ is compactly embedded in $L^2(\R^2)$.

Let us consider $D\coloneqq\{u\in \mathcal{V}_{h,\A} : \Vert u \Vert_{\mathcal{V}_{h,\A}} \leq 1\}$ and prove its precompactness in $L^2(\R^2)$ by means of the Kolmogorov--Riesz theorem (see \cite[Thm.~4.14]{CR21}). We only need to check the following:
\begin{enumerate}[label=\textup{\textbf{(\roman*)}}, ref=\textup{\textbf{\roman*}}]
\item \label{Integrable} For all $\varepsilon>0$, there exists $\omega\subset\subset \R^2$ such that $\int_{\R^2\setminus \omega} |u|^2\, \dd x\leq \varepsilon^2$, for all $u\in D$.
\item \label{Translation} For all $\varepsilon>0$ and for all $\omega \subset\subset \R^2$, there exists $\delta>0$ such that 
\[\int_{\omega} \vert u(x+s)-u(x)\vert^2\, \dd x\leq \varepsilon^2\]
for all $s\in\R^2$ with $|s|\leq \delta$ and for all $u\in D$.
\end{enumerate}
The equi-integrability condition \eqref{Integrable} is satisfied as long as at least one of the assumptions \eqref{Re B to inf}, \eqref{Im B to inf}, or \eqref{Im A to inf} holds. This follows from Assumption~\ref{assumption.1}, Lemma~\ref{Lem Mag Ineqs}, Proposition~\ref{Prop V}\ref{Dense}, and the continuity of the functions considered in assumptions \eqref{Re B to inf}, \eqref{Im B to inf}, and \eqref{Im A to inf}. Below, we provide a detailed proof of \eqref{Integrable} under the assumption \eqref{Re B to inf}. Since the proof is identical under the other assumptions, we will omit those cases.

Assume that $\Re \B$ satisfies \eqref{Re B to inf}. Since $\Re \B$ is continuous, by the multivariate intermediate value theorem \cite[Thm.~1.9.5]{Duistermaat-Kolk04} and the connectedness of the punctured disk in $\R^2$, the condition $\displaystyle\lim_{\vert x \vert \to +\infty} \vert \Re \B(x) \vert = +\infty$ implies that
$$\lim_{\vert x \vert \to +\infty}  \Re \B(x) = +\infty \qquad \text{or} \qquad\lim_{\vert x \vert \to +\infty} -\Re \B(x) = +\infty\,.$$

Without loss of generality, we assume the first possibility. 
By adding a constant $C>0$ such that $\Re\B+C\geq 0$ on $\R^2$ (to apply Fatou's lemma in the following step), and using the first inequality in Lemma \ref{Lem Mag Ineqs}, we obtain
\[ (C+1)\Vert u \Vert_{\mathcal{V}_{h,\A}}^2\geq \int_{\R^2} (\Re \B +C)\vert u \vert^2\, \dd x\,, \qquad \forall u \in C_{c}^{\infty}(\R^2)\,.\]
To extend this inequality to the space $\mathcal{V}_{h,\A}$, we utilise the density of $C_{c}^{\infty}(\R^2)$ in $\mathcal{V}_{h,\A}$. More precisely, let $u\in \mathcal{V}_{h,\A}$, there exists a sequence $\left(u_{n}\right)_{n \in \N} \subset  C_{c}^{\infty}(\R^2)$ such that $u_{n}\xrightarrow{n\to+\infty} u$ in $\mathcal{V}_{h,\A}$. By considering a subsequence, still denoted by $u_{n}$, such that $u_{n}(x)\xrightarrow{n\to+\infty} u(x)$ for almost every $x\in \R^2$, Fatou's lemma yields that
\begin{equation*}
\begin{aligned}
(C+1) \Vert u \Vert_{\mathcal{V}_{h,\A}}^2= \lim_{n\to+\infty} (C+1) \Vert u_{n} \Vert_{\mathcal{V}_{h,\A}} \geq  &\liminf_{n\to+\infty} \int_{\R^2} (\Re \B +C)\vert u_{n} \vert^2\, \dd x\\
\geq &\int_{\R^2} (\Re \B +C)\vert u \vert^2\, \dd x\,, \qquad \forall u\in \mathcal{V}_{\A}\,.
\end{aligned}
\end{equation*}
Given $\varepsilon>0$, using this estimate and the unboundedness of $\Re \B$ at infinity, there exists a constant $R>0$ such that
\[ \int_{\vert x \vert> R} \vert u \vert^2 \, \dd x < \varepsilon\,, \qquad \forall u \in D\,. \]
Let us now consider \eqref{Translation}. Let $\varepsilon>0$ and $\omega\subset\subset \R^2$, we consider a function $\chi \in C_{c}^{\infty}(\R^2)$ and $\chi=1$ in some neighbourhood of $\overline{\omega}$. For all $u\in D$,
\[ -i h \nabla(\chi u)=\Re \mathbf{A} \chi u+(-i h \nabla\chi)u +\chi(-i h \nabla-\Re \mathbf{A})u\in L^2(\R^2)\,,\]
where we used the fact that $\Re \mathbf{A}\in L^\infty_{\mathrm{loc}}(\R^2,\R^2)$.
There exists $C>0$ such that, for all $u\in D$,
\[ \Vert  \chi u  \Vert_{H^1(\R^2)}\leq C\,.\]
Then, we notice that, for $|s|$ small enough such that $\chi(x+s)=1$ on $\omega$, we have, for all $u\in D$,
\begin{align*}
	\int_{\omega}|u(x+s)-u(x)|^2\,\dd x =&\int_{\omega} \vert(\chi u)(x+s)-(\chi u)(x)|^2\,\dd x\\
	\leq &\int_{\R^2}|(\chi u)(x+s)-(\chi u)(x)|^2\,\dd x\leq C^2 |s|^2\,,
\end{align*}
where we used  \cite[Prop.~2.94]{CR21}.
\end{proof}
%

\section{WKB construction of pseudomodes}\label{Sec WKB}
%
This section is concerned with task~(III)
about the construction of semiclassical pseudomodes. 
In particular, we establish Theorem~\ref{Theorem exponential}.
Throughout this section, we assume that $\B\in C^{\infty}(\R^2)$.

\begin{definition}
 We say that $\B$ is real-analytic at $x^{0}$, when, in the neighbourhood of $x^0$, $\mathbf{B}$ is the sum of a converging series:
	\[ \B(x_{1},x_{2}) = \sum_{m,n\geq 0} b_{m n} (x_1-x_1^0)^{m}(x_2-x_2^0)^{n},\qquad b_{mn}\in \C\,.\]
\end{definition}

The main technical result of this section is the following theorem.
\begin{theorem}\label{Theorem pseudomode}
	Let $x^{0}\in \Gamma$ and assume that $\B$ is real-analytic at $x^{0}$. Then, there exist 
	\begin{enumerate}[\rm i)]
		\item a neighbourhood $\mathcal{U}$ of $x^0$ in $\R^2$;
		\item a real-analytic function $P$ on $\mathcal{U}$ satisfying
		\begin{equation}\label{Phase S}
			\mathrm{Re}\, P(x) = Q(x_1-x_1^0,x_2-x_2^0) + \mathcal{O}\left(\vert x-x^{0} \vert^3\right),
		\end{equation}
		where $Q(u,v):= Q_1(x^{0})u^2 -2Q_2(x^{0}) uv + Q_3(x^{0}) v^2$ is a positive definite quadratic form on $\R^2$ with $Q_1,Q_2, Q_3$ defined as in \eqref{Q123};
		\item a sequence of real-analytic functions $(a_{j})_{j\in \N}$ on $\mathcal{U}$ with $a_0(x^0)=1$ and $a_{j}(x^0)=0$ for $j\geq 1$;
	\end{enumerate}
	such that, for all $N\in \N$,
	\begin{equation}
		e^{P/h}\left( \mathscr{L}_{h,\A} - h\B(x^{0})\right)\left(e^{-P/h}\sum_{j=0}^{N} h^{j}a_j\right)=\mathcal{O}\left(h^{N+2}\right)
	\end{equation}
	locally uniformly on $\mathcal{U}$.
\end{theorem}

After a translation, we can assume that $x^{0}=0$.
\begin{notation}[Complexification of a real-analytic function]\label{Complexify Not}
Assume that $a$ is real-analytic near the point $0\in \R^2$. 
We denote by $\widetilde{a}$ the function defined near $0\in \C^2$ by
\[ \widetilde{a}(z,w) \coloneqq a\left(\frac{z+w}{2},\frac{z-w}{2i} \right)\,.\]
Note that 
\begin{equation}\label{Change of variable}
\widetilde{a}(z,\overline{z})=a(\mathrm{Re}\, z, \mathrm{Im}\, z)\,, \qquad \partial_{z} \widetilde{a} = \widetilde{\partial_{z} a}\,,\qquad  \partial_{w} \widetilde{a} = \widetilde{\partial_{\overline{z}} a}\,.
\end{equation}
\end{notation} 
 
\subsection{A choice of the magnetic potential}\label{Choice MP}
\begin{lemma}\label{Poisson Sol}
There exists a real-analytic and complex-valued function $\varphi$ in a neighbourhood $\Omega$ of $0$ such that
\[ \Delta \varphi = \B\,,\qquad\varphi(x_{1},x_{2}) = \frac{\B(0)}{4}\left( x_{1}^2+x_2^2\right)+ \mathcal{O}\left(|x|^3\right)\,. \]
\end{lemma}
\begin{proof}
By considering the complexification of $\B$ in the neighbourhood of $0$
\[\widetilde{\B}(z,w)=\sum_{(\alpha,\beta)\in \N^2} a_{\alpha,\beta} z^{\alpha}w^{\beta}\,,\]
with $a_{0,0}=\B(0)$, we introduce the power series
\[ \widetilde{\varphi}(z,w)=\frac{1}{4}\sum_{(\alpha,\beta)\in \N^2}\frac{ a_{\alpha,\beta} }{(\alpha+1)(\beta+1)}z^{\alpha+1}w^{\beta+1}\,.\]
Then we get
\[4\partial_{z}\partial_{w} \widetilde{\varphi}(z,w)=\widetilde{\B}(z,w)\,. \]
The function $z \mapsto \widetilde{\varphi}(z,\overline{z})$ satisfies the required properties. 
\end{proof}
Let $\varphi$ be the function given by Lemma \ref{Poisson Sol}, and define 
$$\mathcal{M}:=\left(-\partial_{x_{2}}\varphi,\partial_{x_{1}}\varphi\right)\,.$$ Then $\mathcal{M}$ satisfies
\[\frac{\partial \mathcal{M}_{2}}{\partial x_{1}} -\frac{\partial \mathcal{M}_{1}}{\partial x_{2}}=\mathbf{B}\qquad \text{on } \Omega\,.\]
By the Poincar\'{e} lemma (say that $\Omega$ is a ball), there exists a function $\theta\in C^{\infty}(\Omega,\C)$ such that
\begin{equation}\label{theta}
 \mathcal{M}=\mathbf{A}+\nabla \theta\qquad \text{on } \Omega\,.
\end{equation}
From this, we obtain
\begin{equation}\label{Relation M and A}
\mathscr{L}_{h,\mathcal{M}}=e^{i\theta/h} \mathscr{L}_{h,\mathbf{A}}e^{-i\theta/h} \qquad \text{on } \Omega\,.
\end{equation}   
 
\subsection{WKB analysis}
In this section, we construct a pseudomode for the operator $\mathscr{L}_{h,\mathcal{M}}$. More precisely, for $N\in \N$, we look for a pseudomode in the form
\[ u_{h}(x)=e^{-S(x)/h}\sum_{j=0}^{N}a_{j}(x)h^{j} \,,\]
attached to a quasi-eigenvalue $\lambda(h)=h\mu$. Here, $S$ and $a_{j}$ are real-analytic functions defined in the neighbourhood of $0\in \R^2$, and $\mu \in  \C$.  

Let us consider the formal conjugated operator acting locally:
\begin{align*}
\mathscr{L}_{h,\mathcal{M}}^{S}:=& \ e^{S/h}\mathscr{L}_{h,\mathcal{M}}e^{-S/h}\\
=& \ (-ih\partial_{1}-\mathcal{M}_{1}+i\partial_{x_1}S)^2 + (-ih\partial_{2}-\mathcal{M}_{2}+i\partial_{x_2}S)^2\\
=& \ E_{0}+hE_{1}+h^2 E_{2} \,,
\end{align*}
where the differential expressions $E_{0}$, $E_{1}$ and $E_{2}$ are given by
\begin{align*}
E_{0} &:= \left(-\mathcal{M}_1+ i \partial_{x_1}S \right)^2 +\left(-\mathcal{M}_2+ i \partial_{x_2}S \right)^2\,,\\
E_{1} &:= \Delta S+2(\nabla S+i\mathcal{M}) \cdot \nabla\,,\\
E_{2} &:=-\Delta\,.
\end{align*}
Then, we have
\begin{align*}
e^{S/h}\left(\mathscr{L}_{h,\mathcal{M}}-\lambda(h)\right)u_{h}(x)&=\left[E_{0}+h(E_{1}-\mu)+h^2 E_{2}\right]\sum_{j=0}^{N}a_{j}(x)h^{j}= \sum_{j=0}^{N+2}\phi_{j}(x)h^{j}\,,
\end{align*}
where the functions $\phi_{j}$ are explicitly given by
\begin{align}
  &h^0 : &E_0 a_0 &=: \phi_{0}\,,\nonumber\\
  &h^1 : &E_0 a_1 + \left(E_1 -\mu \right)a_0 &=:\phi_{1}\,,\nonumber\\
  &h^2 : &E_0 a_2 + \left(E_1 -\mu \right)a_1 +E_{2} a_0 &=: \phi_{2}\,,\nonumber\\
  &\vdots &&\label{eq.system}\\
  &h^{N} : &E_0 a_{N} + \left(E_1 -\mu \right)a_{N-1}+E_{2}a_{N-2} &=:\phi_{N} \,,\nonumber\\
  &h^{N+1} : & \left(E_{1}-\mu\right)a_{N}+E_{2}a_{N-1}&=:\phi_{N+1}\,,\nonumber
\end{align}
and the last function $\phi_{N+2}$ is
\begin{equation*}
\begin{aligned}
h^{N+2}: E_{2}a_{N}&=:\phi_{N+2}\,.
\end{aligned}
\end{equation*} 
 
\subsubsection{The eikonal equation}
Let us find $S$ such that $E_{0}=0$, \emph{i.e.},
\[ (-\mathcal{M}_{1}+i\partial_{x_1}S)^2+(-\mathcal{M}_2+i\partial_{x_2}S)^2=0\,.\]
It is equivalent to the equation
\[\left(i\partial_{x_1}S -\partial_{x_2}S-\mathcal{M}_1-i\mathcal{M}_2\right)\left(i\partial_{x_1}S +\partial_{x_2}S-\mathcal{M}_1+i\mathcal{M}_2\right)=0\,. \]
Let us choose $S$ such that
\[ i\partial_{x_1}S -\partial_{x_2}S-\mathcal{M}_1-i\mathcal{M}_2 =0 \,, \]
the other choice leading to a phase that does not provide us with an exponentially decaying quasimode (as in the selfadjoint situation in \cite{BR-20}).
 
Then we have
\[ 2\partial_{\overline{z}} S = \mathcal{M}_2-i\mathcal{M}_1 =2\partial_{\overline{z}} \varphi\,.\]
In particular, $S-\varphi$ is holomorphic. By using Notation \ref{Complexify Not}, this suggests taking
\[ \widetilde{S}(z,w)= \widetilde{\varphi}(z,w)+ f(z)\,,\]
where $f(z)$ is a holomorphic function (in a neighbourhood of $0\in \R^2$) to be determined later in the neighbourhood of $0\in \C$. Note that $\Delta S=\mathbf{B}$. 
 
\subsubsection{Towards the transport equations}
Now, we consider the operator $E_{1}$.

By using the expression of $\mathcal{M}$ and the choice of $\widetilde{S}$, we have
\begin{align*}
&(\nabla S+i\mathcal{M})\cdot \nabla=\lefteqn{\left( \partial_{x_{1}} S+i\mathcal{M}_{1}\right)\partial_{x_{1}}+ \left( \partial_{x_{2}} S+i\mathcal{M}_{2}\right)\partial_{x_{2}}}\\
=& \left(\partial_{x_{1}} \varphi-i\partial_{x_2}\varphi+f'(z)\right)\partial_{x_{1}}+\left(\partial_{x_{2}} \varphi+i\partial_{x_1}\varphi+if'(z)\right)\partial_{x_{2}}\\ 
=&2\left(2\partial_{z} \varphi+f'(z)\right)\partial_{\overline{z}}\,.
\end{align*}
Thus, for any real-analytic function $a$ near the point $x^{0}$, we have
\[ \widetilde{E_{1}a} =\left[4\left(2\partial_{z} \widetilde{\varphi}+f'(z)\right)\partial_{w}  +\widetilde{\B}\right]\widetilde{a}\,.\]
Note also that
\[ \widetilde{E_{2}a}=-4\partial_{z}\partial_{w} \widetilde{a}\,.\] 
From \eqref{eq.system}, we are led to the the system of the transport equations
\begin{align*}
  &h^1 : &&\left[4\left(2\partial_{z} \widetilde{\varphi}+f'(z) \right)\partial_{w}+\widetilde{\B} -\mu \right]\widetilde{a}_{0} = 0\,,\\
  &h^2 : &&\left[4\left(2\partial_{z} \widetilde{\varphi}+f'(z) \right)\partial_{w}+\widetilde{\B} -\mu \right]\widetilde{a}_{1} = 4\partial_{z}\partial_{w} \widetilde{a}_{0}\,,\\
  &\vdots &&\\
  &h^{N+1} : &&\left[4\left(2\partial_{z} \widetilde{\varphi}+f'(z) \right)\partial_{w}+\widetilde{\B} -\mu \right]\widetilde{a}_{N}= 4\partial_{z}\partial_{w} \widetilde{a}_{N-1}\,.
\end{align*} 
 
\subsubsection{Choosing $f$ and determining $S$}
Since $0\in\Gamma$, we have $\partial_{\overline{z}} \B(0)\neq 0$ and thus
\[ \partial_{w}\widetilde{\B}(0) =\widetilde{\partial_{\overline{z}} \B}(0)=\partial_{\overline{z}} \B(0)\neq 0\,.\]
In virtue of the holomorphic implicit function theorem, there exists a unique holomorphic function $w$, in the neigborhood of $0\in \C$, such that
\begin{equation}\label{function w}
w(0)= 0, \qquad \widetilde{\B}(z,w(z))=\B(0)\,.
\end{equation}
In particular,
\begin{equation}\label{eq.w'0}
	w'(0)=-\frac{\partial_{z} \widetilde{\B}}{\partial_{w} \widetilde{\B}}(0)=-\frac{\partial_{z} \B}{\partial_{\overline{z}} \B}(0)\,.
	\end{equation}
	\begin{remark}
Note that the function $w$ solves the same effective equation as in  \cite{BR-20}, but that we are \emph{not} in the case of a magnetic well.		
		\end{remark}
In order to solve the above transport equations, we will choose a function $f(z)$ such that 
\[ 2\partial_{z} \widetilde{\varphi}(z,w(z))+f'(z) =0\,.\] 
 
\begin{lemma}\label{Lemma f}
Let $\varphi$ be given in Lemma \ref{Poisson Sol}, $w$ be the holomorphic function given in \eqref{function w} and $\theta$ given in \eqref{theta}. Letting
\begin{equation}\label{eq.f}
f(z) := -\int_{[0,z]} 2\partial_{z} \widetilde{\varphi}(\zeta,w(\zeta)) \, \dd \zeta+ \Im \theta(0)\,,
\end{equation}
we have
\[f(0)=\Im \theta(0),\qquad f'(0)=0, \qquad f''(0) = \frac{\B(0)}{2} \frac{\partial_{z}\B}{\partial_{\overline{z}}\B}(0)\,.\]
In particular, 
\[\Re S(x)= \Im \theta(0)+ \tilde Q_1 x_1^2 -2\tilde Q_2x_1 x_2+ \tilde Q_3 x_2^2+ \mathcal{O}(\vert x \vert^3)\,,\]
where 
\begin{align*}
			&\tilde Q_1 = \frac{1}{4} \textup{Re}\,\left[ \B(0)\left(1+\frac{\partial_{z} \B}{\partial_{\overline{z}} \B}(0)\right)\right],\qquad \tilde Q_2 = \frac{1}{4} \mathrm{Im}\, \left[ \B(0)\frac{\partial_{z} \B}{\partial_{\overline{z}} \B}(0)\right],\\
			&\tilde Q_3 =  \frac{1}{4} \textup{Re}\,\left[ \B(0)\left(1-\frac{\partial_{z} \B}{\partial_{\overline{z}} \B}(0)\right)\right].
		\end{align*}
\end{lemma} 
\begin{proof}
A straightforward computation gives
\[ f''(0)=-2\partial_{zz}^{2}\widetilde{\varphi}(0)-2w'(0)\partial^{2}_{zw} \widetilde{\varphi}(0)\,.\]
Due to Lemma \ref{Poisson Sol}, we have 
\[\partial_{zz}^2\widetilde{\varphi}(0)=0\,,\quad \partial_{zw}^{2} \widetilde{\varphi}(0)=\frac{\B(0)}{4}\,,\]
and we get the value of $f''(0)$ by using \eqref{eq.w'0}.

Since $S(x_{1},x_{2})=\varphi(x_{1},x_{2})+f(x_{1}+ix_{2})$, we can write
\begin{align*}
S(x_1,x_2) &=\Im \theta(0)+ \frac{\B(0)}{4} \left[x_1^2+x_2^2+\frac{\partial_{z}\B(0)}{\partial_{\overline{z}}\B(0)} \left(x_1+ix_2\right)^2 \right]+\mathcal{O}(\vert x \vert^3)\,,
\end{align*}
so that
\begin{align*}
\textup{Re}\, S(x) =&\ \Im \theta(0)+ \frac{1}{4} \textup{Re}\,\left[ \B(0)\left(1+\frac{\partial_{z} \B(0)}{\partial_{\overline{z}} \B(0)}\right)\right]x_1^2+\frac{1}{4} \textup{Re}\,\left[ \B(0)\left(1-\frac{\partial_{z} \B(0)}{\partial_{\overline{z}} \B(0)}\right)\right]x_2^2\\
&-\frac{1}{2} \mathrm{Im}\, \left[ \B(0)\frac{\partial_{z} \B(0)}{\partial_{\overline{z}} \B(0)}\right]x_1x_2+\mathcal{O}(\vert x \vert^3)\\
=& \ \Im \theta(0)+\tilde Q_1x_1^2 -2 \tilde Q_2 x_1x_2+\tilde Q_3 x_2^2+ \mathcal{O}(\vert x \vert^3)\,.
\end{align*}
\end{proof}

\subsubsection{Solving the first transport equation}
With the choice \eqref{eq.f}, we can write
\begin{equation}\label{First Transport Eq}
8\left[\partial_{z}\widetilde{\varphi}(z,w)-\partial_{z}\widetilde{\varphi}(z,w(z))\right]\partial_{w}\widetilde{a}_{0}(z,w) + \left(\widetilde{\B}(z,w)-\mu\right)\widetilde{a}_0(z,w) = 0\,.
\end{equation}
Taking $w=w(z)$ and using \eqref{function w}, we see that \eqref{First Transport Eq} has a holomorphic solution $\widetilde{a}_{0}$ such that $\widetilde{a}_{0}(0)\neq 0$ if and only if
\[\mu=\B(0)\,. \]
Let us explain this. The Taylor formula gives
\[
\begin{aligned}
  \partial_{z}\widetilde{\varphi}(z,w)-\partial_{z}\widetilde{\varphi}(z,w(z)) 
  &=(w-w(z))V(z,w)\,, 
  \\
  \widetilde{\B}(z,w)-\widetilde{\B}(z,w(z)) &= (w-w(z))F(z,w)\,,
\end{aligned}   
\]
where
\[V(z,w)=\int_{0}^1 \widetilde{\B}(z,w(z)+t(w-w(z)) \, \dd t,\quad F(z,w)=\int_{0}^1 \partial_{w}\widetilde{\B}(z,w(z)+t(w-w(z)) \, \dd t\,.
\]
Now suppose that \eqref{First Transport Eq} has a holomorphic solution $\widetilde{a}_{0}$ such that $\widetilde{a}_{0}(0)\neq 0$. Substituting $(z,w)=(0,0)$ in \eqref{First Transport Eq} and using $w(0)=0$ from \eqref{function w}, we obtain $\mu=\widetilde{\B}(0,0)=\B(0)$. Notice that $V(0)=\B(0)\neq 0$, so that near $0\in \C^2$, we have $V(z,w)\neq 0$. By using $\mu=\B(0)$, the first transport equation becomes
\[ \partial_{w}\widetilde{a}_{0}(z,w) + \frac{1}{8}\frac{F(z,w)}{V(z,w)} \widetilde{a}_{0}(z,w)=0\,.\]
Therefore, we have
\[ \widetilde{a}_{0}(z,w) = \mathcal{A}_{0}(z) J(z,w)\,,\quad J(z,w) \coloneqq\exp\left(-\int_{[w(z),w]}\frac{1}{8}\frac{F(z,u)}{V(z,u)}  \, \dd u\right)\,,\]
where $\mathcal{A}_{0}(z)$ is a holomorphic function to be determined such that $\mathcal{A}_{0}(0)=1$.
\subsubsection{Solving the second transport equation}
Let us now consider the second transport equation
\begin{equation}\label{Second Transport Eq}
(w-w(z))\left[8 V(z,w)\partial_{w} + F(z,w)\right]\widetilde{a}_{1}(z,w) =4\partial_{z}\partial_{w}  \widetilde{a}_{0}(z,w)\,.
\end{equation}
Letting $w=w(z)$, we necessarily get that
\[4\partial_{z}\partial_{w}\widetilde{a}_{0}(z,w(z))=0\,,\]
which means that
\[\partial_{w}J(z,w(z)) \mathcal{A}_{0}'(z)+ \partial_{z}\partial_{w}J(z,w(z))\mathcal{A}_{0}(z)= 0\,.\]
 This allows us to determine $\mathcal{A}_{0}$ of the previous step. From the definition of $J$, we have
 \[ \partial_{w}J(z,w(z)) = -\frac{1}{8}\frac{F(z,w(z))}{V(z,w(z))}\,,\]
and thus
 \[ \partial_{w}J(0,w(0)) = - \frac{1}{8} \frac{\partial_{w}\widetilde{\B}}{\widetilde{\B}}(0)=- \frac{1}{8} \frac{\partial_{\overline{z}}\B}{\B}(0)  \neq 0\,.\]
This leads to the choice
\[ \mathcal{A}_{0}(z)= \exp\left( -\int_{[0,z]} \frac{\partial_{z}\partial_{w} J}{\partial_{w} J}(u,w(u))\, \dd u\right)\,,\]
which is not allowed in the case of a magnetic well in the self-adjoint case \cite{BR-20}, since  $\frac{\partial_{z}\partial_{w} J}{\partial_{w} J}(z,w(z)) \approx \frac{1}{z}$ as $z\to 0$ (to be compared with \cite[Sec. 3.4]{BR-20}).

In particular,
\[\widetilde{a}_{0}(z,w)=\exp\left( -\int_{[0,z]} \frac{\partial_{z}\partial_{w} J}{\partial_{w} J}(u,w(u))\, \dd u\right)\exp\left(-\int_{[w(z),w]} \frac{1}{8}\frac{F(z,u)}{V(z,u)}  \, \dd u\right)\,.\]
With this choice, \eqref{Second Transport Eq} becomes
\[ \partial_{w}\widetilde{a}_{1}(z,w) + \frac{1}{8}\frac{F(z,w)}{V(z,w)}\widetilde{a}_{1}(z,w) =\frac{1}{2}\frac{T_0(z,w)}{V(z,w)}\,,\]
with
\[\displaystyle T_{0}(z,w)=\int_{0}^1 \partial_{w}^{2}\partial_{z}\widetilde{a}_{0}(z,w(z)+t(w-w(z)))\, \dd t\,.\]
We have
\[ \widetilde{a}_{1}(z,w) = J(z,w) \int_{[w(z),w]}\frac{T_0(z,u)}{2J(z,u)V(z,u)} \,\dd u+ \mathcal{A}_{1}(z) J(z,w),\]
where $\mathcal{A}_{1}(z)$ is a holomorphic function to be determined with $\mathcal{A}_1(0)=0$.
 
\subsubsection{Induction}
By induction, the solution of the $j+1$-th transport equation can be written as
\[\widetilde{a}_{j+1}(z,w) = J(z,w)\int_{[w(z),w]} \frac{T_{j}(z,u)}{2J(z,u)V(z,u)} \,\dd u+ \mathcal{A}_{j+1}(z) J(z,w)\,,\]
with
$$\displaystyle T_{j}(z,w)=\int_{0}^1 \partial_{w}^{2}\partial_{z}\widetilde{a}_{j}(z,w(z)+t(w-w(z)))\, \dd t$$
and $\mathcal{A}_{j+1}$ is determined by the constraint (coming from the $j+2$-th equation):
\begin{equation}
4\partial_{z}\partial_{w}\widetilde{a}_{j+1}(z,w(z))=0\,. 
\end{equation}
We have
\[\partial_{w}\widetilde{a}_{j+1}(z,w) = \partial_{w}J(z,w) \int_{[w(z),w]} \frac{T_{j}(z,u)}{2J(z,u)V(z,u)} \,\dd u+  \frac{T_{j}(z,w)}{2V(z,w)}+\mathcal{A}_{j+1}(z)\partial_{w}J(z,w)\,,
\]
so that
\begin{align*}
&\partial_{z}\partial_{w}\widetilde{a}_{j+1}(z,w(z)) \\
= &\left[-\partial_{w}J\,w'(z)  \frac{T_{j}}{2V}+\partial_{z}J \frac{T_{j}}{2V}+\partial_{z}\left(\frac{T_{j}}{2V} \right)+\mathcal{A}_{j+1}'\partial_{w}J+\mathcal{A}_{j+1}\partial_{zw}^{2}J\right](z,w(z))\,.
\end{align*}
Since $J(z,w(z))=1$, we have $\partial_{z}J(z,w(z))=-w'(z)\partial_{w}J(z,w(z))$.
Therefore, $\mathcal{A}_{j+1}$ satisfies the equation
\[ \partial_{w}J(z,w(z))\mathcal{A}_{j+1}'(z)+\partial_{zw}^{2}J(z,w(z))\mathcal{A}_{j+1}(z)=G_j(z)\,,\]
where
\[G_j(z)\coloneqq-\left[\partial_{z}J\frac{T_{j}}{V}+\partial_{z}\left(\frac{T_{j}}{2V} \right)\right](z,w(z))\,.\]
Thus,
\begin{equation*}
\mathcal{A}_{j+1}(z)=\mathcal{A}_{0}(z)\int_{[0,z]}\frac{G_{j}(u)}{\mathcal{A}_{0}(u)}\,\dd u\,.
\end{equation*}
Therefore, we have, for all $j\geq 0$,
\begin{equation}\label{Transport Solution j+1}
\widetilde{a}_{j+1}(z,w) = J(z,w) \int_{[w(z),w]} \frac{T_{j}(z,u)}{2J(z,u)V(z,u)} \,\dd u+ \widetilde{a}_{0}(z,w)\int_{[0,z]}\frac{G_{j}(u)}{\mathcal{A}_{0}(u)}\,\dd u\,,
\end{equation}
where
\[T_{j}(z,w)=\int_{0}^1 \partial_{w}^{2}\partial_{z}\widetilde{a}_{j}(z,w(z)+t(w-w(z)))\, \dd t\,,\]
and
\[G_{j}(u)=\left[\left(\frac{\partial_{z}J}{\B(0)} +\frac{1}{2}\partial_{z}\left(\frac{1}{V}\right)\right)\partial_{w}^2\partial_{z}\widetilde{a}_{j}+\frac{1}{2\B(0)}\partial_{w}^2\partial_{z}^2 \widetilde{a}_{j}+\frac{w'(u)}{4\B(0)}\partial_{w}^3\partial_{z}\widetilde{a}_{j}\right](u,w(u))\,.\]

\subsection{Proof of Theorem \ref{Theorem pseudomode}}
We have constructed holomorphic functions $\widetilde{S}(z,w)$ and $\left(\widetilde{a}_{j}(z,w) \right)_{j\in \N_{0}}$ which are defined in a neighbourhood of $0\in \C^2$. Now we take $w=\overline{z}$ and recall Notation \ref{Complexify Not}. For all $N\geq 0$, we have
 \begin{equation*}
e^{S/h}\left( \mathscr{L}_{h,\mathcal{M}} - h\mu\right)\left(e^{-S/h}\sum_{j=0}^{N} h^{j}a_j\right)= (-\Delta a_{N}) h^{N+2}\,.
\end{equation*}
From \eqref{Relation M and A}, we have
\begin{equation}\label{Conclude}
e^{P/h}\left( \mathscr{L}_{h,\A} - h\mu\right)\left(e^{-P/h}\sum_{j=0}^{N} h^{j}a_j\right)= (-\Delta a_{N}) h^{N+2}\,,\quad P:=S+i\theta\,.
\end{equation}
From the definitions of $\theta$ and $\mathcal{M}$, we have $\nabla \theta(0)= \mathcal{M}(0)-\A(0)=-\A(0)$ and thus 
$\nabla \Im \theta(0)=-\Im \A(0)= 0$. Notice that
\begin{align*}
\Im \partial_{1}^2 \theta(0) 
&= \Im \left( -\partial_{12}^{2} \varphi(0)-\partial_{1} A_{1}(0) \right) = -\partial_{1} \Im A_{1} (0)\,,\\
\Im \partial_{2}^2 \theta(0) 
&= \Im \left( \partial_{12}^{2} \varphi(0)-\partial_{2} A_{2}(0) \right) = -\partial_{2} \Im A_{2} (0)\,,\\
\Im \partial_{12}^2 \theta(0) 
&= \Im \left( \partial_{11}^{2} \varphi(0)-\partial_{1} A_{2}(0) \right)\\
& = \frac{\Im \B(0)}{2}-\partial_{1} \Im A_{2} (0)\\
&=-\frac{1}{2}\left(\partial_{1} \Im A_{2} (0)+\partial_{2} \Im A_{1} (0)\right)\,.
\end{align*}
From Lemma \ref{Lemma f} and using Taylor expansion for $\Im \theta$, we deduce that
\begin{align*}
\Re P(x)= Q_1(0) x_1^2 -2Q_2(0)x_1 x_2+Q_3(0) x_2^2+ \mathcal{O}(\vert x \vert^3)\,.
\end{align*}
This concludes the proof of Theorem \ref{Theorem pseudomode}.
 
\subsection{Proof of Theorem \ref{Theorem exponential}}

\subsubsection{Preliminaries}
\begin{notation}
We will use the following notation for a polydisc
\[ P(0;R_{1},R_{2})\coloneqq\left\{ (z,w)\in \C^2: \vert z\vert <R_{1} \text{ and }\vert w\vert <R_{2}\right\}\,.\]
For $K\subset \C^2$, we write
\[ \Vert \widetilde{a}(z,w) \Vert_{K}\coloneqq\sup_{(z,w)\in K} \vert \widetilde{a}(z,w)\vert\,.\]
When $K=P(0;R_1,R_2)$, we simply write $\|\cdot\|_{K}=\|\cdot\|_{R_1,R_2}$. For $m,n\in \N$, we denote $\llbracket m,n\rrbracket\coloneqq [m,n] \cap \Z$.
\end{notation} 
 
\begin{lemma}\label{lem.Cauchy}
For a holomorphic function $\widetilde{a}(z,w)$ defined in a neighbourhood of $P(0;R_{1},R_{2})$, we have, for all $(z,w)\in P(0;R_{1},R_{2})$,
\[	\left\vert \partial_{w}^2\partial_{z} \widetilde{a}(z,w)\right\vert\leq  \frac{2R_{1}R_{2} \Vert \widetilde{a} \Vert_{R_{1},R_{2}}}{\left( R_{1}-\vert z\vert \right)^2\left( R_{2}-\vert w\vert \right)^3}\]
and
\begin{equation*}
	\left\vert \partial_{w}^2\partial_{z}^2 \widetilde{a}(z,w)\right\vert \leq\frac{4R_{1}R_{2} \Vert \widetilde{a} \Vert_{R_{1},R_{2}}}{\left( R_{1}-\vert z\vert \right)^3\left( R_{2}-\vert w\vert \right)^3}\,,\quad
	\left\vert \partial_{w}^3\partial_{z} \widetilde{a}(z,w)\right\vert \leq\frac{6R_{1}R_{2} \Vert \widetilde{a} \Vert_{R_{1},R_{2}}}{\left( R_{1}-\vert z\vert \right)^2\left( R_{2}-\vert w\vert \right)^4}\,.
\end{equation*}	
\end{lemma}
\begin{proof}
For example, the Cauchy formula gives
\begin{equation*}
	\begin{aligned}
		\left\vert \partial_{w}^2\partial_{z} \widetilde{a}(z,w)\right\vert=& \left\vert \frac{2}{(2\pi i)^2} \int_{\vert z\vert=R_{1}} \int_{\vert w\vert=R_{2}}\frac{\widetilde{a}(\xi_{1},\xi_{2})}{(\xi_{1}-z)^2(\xi_{2}-w)^3} \, \dd \xi_{1} \dd \xi_{2}\right\vert\\
		\leq & \frac{2R_{1}R_{2} \Vert \widetilde{a} \Vert_{R_{1},R_{2}}}{\left( R_{1}-\vert z\vert \right)^2\left( R_{2}-\vert w\vert \right)^3}\,.
	\end{aligned}
\end{equation*}

\end{proof} 
 
\begin{lemma}\label{lem.boundaj}
Let $(a_{j})_{j\in \N}$ be the real-analytic sequence given in Theorem \ref{Theorem pseudomode}, then there exist constants $m>0$, $\delta_{0}>0$ such that, for all $j\geq 0$ and for all $x\in D(x^{0},\delta_{0})$,
\begin{equation}\label{Transport Est}
\left\vert a_{j}(x)\right\vert \leq m^{j+1} j^{7j},\qquad \left\vert \nabla a_{j}(x)\right\vert \leq m^{j+1} j^{7j},\qquad \left\vert \Delta a_{j}(x)\right\vert \leq m^{j+1} j^{7j}\,.
\end{equation}
\end{lemma}
\begin{proof}
 Let us prove that there exists $m>0$ such that, for all $j\geq 0$ and all $(z,w)$ in a neighbourhood of $0\in\C^2$,
\begin{equation}\label{Estimate Transport}
\left\vert \widetilde{a}_{j}(z,w) \right\vert \leq m^{j+1} j^{7j}\,.
\end{equation}
The key to get the growth control is the recursion relation \eqref{Transport Solution j+1}: we see that $\widetilde{a}_{j+1}$ is related to derivatives $\partial_{w}^2\partial_{z}\widetilde{a}_{j}$, $\partial_{w}^2\partial_{z}^2\widetilde{a}_{j}$, and $\partial_{w}^3\partial_{z}\widetilde{a}_{j}$. 

Consider $C, R_1,R_2>0$ such that, for all $z\in D(0,R_1)$,
\begin{equation}\label{Bound for w}
\vert w(z) \vert \leq C |z|<R_{2}\,.
\end{equation}
For all $(z,w)\in P\left(0;R_{1},R_{2}\right)$ and for all $t\in [0,1]$, we have
\begin{equation}\label{Insert w}
\vert w(z)+t(w-w(z))\vert <\max(C|z|,|w|)<R_2\,.
\end{equation}
Then, Lemma \ref{lem.Cauchy} yields that, for all $(z,w)\in P\left(0;R_{1},R_{2}\right)$,
\begin{align*}
\left\vert T_{j}(z,w)\right\vert \leq &\int_{0}^{1} \vert \partial_{w}^2\partial_{z} \widetilde{a}_{j}(z,w(z)+t(w-w(z))\vert\, \dd t\\
\leq &\int_{0}^{1} \frac{2R_{1}R_{2} \Vert \widetilde{a}_{j} \Vert_{R_{1},R_{2}}}{\left( R_{1}-|z|\right)^2\left( R_{2}-|w(z)+t(w-w(z))|\right)^3}\, \dd t\\
\leq & \frac{2R_{1} R_{2} \Vert \widetilde{a}_{j} \Vert_{R_{1},R_{2}}}{\left( R_{1}-|z|\right)^2\left( R_{2}-\max \{C|z|,|w|\}\right)^3}\,.
\end{align*}
Then, the first term in \eqref{Transport Solution j+1} can be controlled, for all $(z,w)\in P(0;R_{1},R_{2})$,
\begin{equation*}
\begin{aligned}
&\left\vert J(z,w)\int_{[w(z),w]} \frac{T_{j}(z,u)}{2J(z,u)V(z,u)} \,\dd u \right\vert\\
=& \left\vert J(z,w) \int_{0}^{1} \frac{T_{j}(z, w(z)+t(w-w(z)))}{2(JV)(z, w(z)+t(w-w(z)))} (w-w(z))\,\dd t \right\vert\\
\leq &\Vert J\Vert_{R_{1},R_{2}}\left\Vert \frac{1}{JV}\right\Vert_{R_{1},R_{2}}\frac{R_{1} R_{2} \Vert \widetilde{a}_{j} \Vert_{R_{1},R_{2}}}{\left( R_{1}-|z|\right)^2\left( R_{2}-\max \{C|z|,|w|\}\right)^3} |w-w(z)|\\
\leq &\Vert J\Vert_{R_{1},R_{2}}\left\Vert \frac{1}{JV}\right\Vert_{R_{1},R_{2}}\frac{2R_{1} R_{2}^2 \Vert \widetilde{a}_{j} \Vert_{R_{1},R_{2}}}{\left( R_{1}-|z|\right)^2\left( R_{2}-\max \{C|z|,|w|\}\right)^3}\,.
\end{aligned}
\end{equation*}
For the second term in \eqref{Transport Solution j+1}, we have, for all $(z,w)\in P\left(0;R_{1},R_{2}\right)$,
\begin{equation*}
\begin{aligned}
&\left\vert \widetilde{a}_{0}(z,w)\int_{[0,z]} \frac{G_{j}(u)}{\mathcal{A}_{0}(u)}\, \dd u \right\vert \\
\leq &|z|\left\Vert \widetilde{a}_{0} \right\Vert_{R_{1},R_{2}} \left\Vert \frac{1}{\mathcal{A}_{0}}\right\Vert_{R_{1}}\\
&\times \int_{0}^{1} \left\vert \left[\left(\frac{\partial_{z}J}{\B(0)} +\frac{1}{2}\partial_{z}\left(\frac{1}{V}\right)\right)\partial_{w}^2\partial_{z}\widetilde{a}_{j}+\frac{1}{2\B(0)}\partial_{w}^2\partial_{z}^2 \widetilde{a}_{j}+\frac{w'(tz)}{4\B(0)}\partial_{w}^3\partial_{z}\widetilde{a}_{j}\right](tz,w(tz)) \right|\, \dd t\\
\leq &|z|\Vert \widetilde{a}_{0} \Vert_{R_{1},R_{2}}\left\Vert \frac{1}{\mathcal{A}_{0}}\right\Vert_{R_{1}}\left[\left(\frac{\Vert \partial_{z}J \Vert_{R_{1},R_{2}}}{\vert \B(0) \vert} +\frac{1}{2}\left\Vert\partial_{z}\left(\frac{1}{V}\right)\right\Vert_{R_{1},R_{2}}\right)\int_{0}^{1} \left\vert \partial_{w}^2\partial_{z} \tilde{a}_{j} (tz,w(tz))\right\vert\, \dd t\right.\\
&\left.+\frac{1}{2|\B(0)|} \int_{0}^{1} \left\vert \partial_{w}^2\partial_{z}^2 \tilde{a}_{j} (tz,w(tz))\right\vert \, \dd t+ \frac{\Vert w' \Vert_{R_{1}}}{4|\B(0)|} \int_{0}^{1} \left\vert \partial_{w}^3\partial_{z} \tilde{a}_{j} (tz,w(tz))\right\vert \, \dd t\right]\,.
\end{aligned}
\end{equation*}
From Lemma \ref{lem.Cauchy}, we have, for all $(z,w)\in P\left(0;R_{1},R_{2}\right)$,
\begin{align*}
\int_{0}^{1} \left\vert \partial_{w}^2\partial_{z} \tilde{a}_{j} (tz,w(tz))\right\vert |z|\, \dd t \leq \frac{2R_{1}^2 R_{2} \left\Vert \widetilde{a}_{j}\right\Vert_{R_{1},R_{2}}}{\left(R_{1}-|z|\right)^2\left(R_{2}-C|z|\right)^2}\,,
\end{align*}
\begin{equation*}
\int_{0}^{1} \left\vert \partial_{w}^2\partial_{z}^2 \tilde{a}_{j} (tz,w(tz))\right\vert |z|\, \dd t\leq \frac{4R_{1}^2 R_{2} \left\Vert \widetilde{a}_{j}\right\Vert_{R_{1},R_{2}}}{\left(R_{1}-|z|\right)^3\left(R_{2}-C|z|\right)^3}\,,
\end{equation*}
and
\begin{equation*}
\int_{0}^{1} \left\vert \partial_{w}^3\partial_{z} \tilde{a}_{j} (tz,w(tz))\right\vert |z|\, \dd t\leq \frac{6R_{1}^2 R_{2} \left\Vert \widetilde{a}_{j}\right\Vert_{R_{1},R_{2}}}{\left(R_{1}-|z|\right)^2\left(R_{2}-C|z|\right)^4}\,.
\end{equation*}
Let us consider $R_{1}^{0}, R_{2}^{0}, C_1, C>0$ such that, for all $R_{1}\leq 2 R_{1}^{0}, R_{2}\leq 2 R_{2}^{0}$,
\[ \Vert J\Vert_{R_{1},R_{2}}+\left\Vert \frac{1}{JV}\right\Vert_{R_{1},R_{2}}+\Vert \widetilde{a}_{0} \Vert_{R_{1},R_{2}}+\left\Vert \frac{1}{\mathcal{A}_{0}}\right\Vert_{R_{1}}+ \Vert \partial_{z}J \Vert_{R_{1},R_{2}}+ \left\Vert\partial_{z}\left(\frac{1}{V}\right)\right\Vert_{R_{1},R_{2}}+\Vert w' \Vert_{R_{1}}\leq C_{1}\,,\]
and, for all $z\in D(0, 2R_{1}^{0})$,
\begin{equation}\label{Assum R1R2}
	|w(z)|\leq C |z|< 2R_{2}^{0}\,.
\end{equation}
From \eqref{Transport Solution j+1}, there exists a constant $C_{2}>0$ (depending only on $R_{1}^{0},R_{2}^{0}$) such that, for all $\ell\geq 0$ and for all $(z,w)\in P\left(0;R_{1},R_{2}\right)$,
\begin{equation}\label{Bound on R}
\vert \widetilde{a}_{\ell+1}(z,w) \vert\leq C_{2}\frac{\Vert \widetilde{a}_{\ell} \Vert_{R_{1},R_{2}}}{\left(R_{1}-\vert z\vert\right)^3\left(R_{2}-\max\{C\vert z\vert, w\}\right)^4}\,.
\end{equation}
We fix $j\in \mathbb{N}$ and let, for all $k\in\llbracket 0,j-1\rrbracket$,
\[ r_{1,k}\coloneqq\left(2-\frac{k}{j}\right)R_{1}^{0}, \qquad r_{2,k}\coloneqq\left(2-\frac{k}{j}\right)R_{2}^{0}\,.\]

Thanks to \eqref{Assum R1R2}, we see that \eqref{Bound for w} holds for $R_{1}=r_{1,k}$ and $R_{2}=r_{2,k}$ with a constant $C$ defined in \eqref{Assum R1R2}. Thus, the estimate \eqref{Bound on R} also holds for all $k\in \llbracket 0,j-1\rrbracket$, \emph{i.e.}, for all $(z,w)\in P(0;r_{1,k},r_{2,k})$,
\begin{equation}\label{Estimate k}
\vert \widetilde{a}_{k+1}(z,w) \vert\leq C_{2}\frac{\Vert \widetilde{a}_{k} \Vert_{r_{1,k},r_{2,k}}}{\left(r_{1,k}-\vert z\vert\right)^3\left(r_{2,k}-\max\{C\vert z\vert, w\}\right)^4}\,.
\end{equation}
This holds in particular for all  $(z,w)\in P(0;r_{1,k+1},r_{2,k+1})$. From the second inequality in \eqref{Assum R1R2}, $CR_{1}^{0}<R_{2}^{0}$, and thus, for $|z|<r_{1,k+1}$, we have
\[ C|z|<C r_{1,k+1}=C\left(2-\frac{k+1}{j}\right)R_{1}^{0}<\left(2-\frac{k+1}{j}\right)R_{2}^{0}=r_{2,k+1}.\]
This shows that, for all $(z,w)\in P(0;r_{1,k+1},r_{2,k+1})$,
\[ \frac{1}{\left(r_{1,k}-\vert z\vert\right)^3\left(r_{2,k}-\max\{C\vert z\vert, w\}\right)^4}\leq  \frac{1}{\left(r_{1,k}-r_{1,k+1}\right)^3\left(r_{2,k}-r_{2,k+1}\right)^4}=\frac{j^7}{(R_{1}^{0})^3(R_{2}^{0})^4}\,.\]
From \eqref{Estimate k} (with $\ell=k$), there exists $C_{3}>0$ such that, for all $k\in \llbracket 0,j-1\rrbracket$, 
\begin{align*}
\Vert \widetilde{a}_{k+1} \Vert_{r_{1,k+1},r_{2,k+1}} \leq C_{3}j^7 \Vert \widetilde{a}_{k} \Vert_{r_{1,k},r_{2,k}}\,.
\end{align*}
Multiplying these estimates, we get
\begin{equation}\label{Estimate for aj}
\Vert \widetilde{a}_{j} \Vert_{R_{1}^{0},R_{2}^{0}}\leq (C_{3}j^7)^{j} \Vert \widetilde{a}_{0} \Vert_{2R_{1}^{0},2R_{2}^{0}}\,.
\end{equation}
Then, the estimate \eqref{Estimate Transport} follows with $m=\max\{C_{3},\Vert \widetilde{a}_{0} \Vert_{2R_{1}^{0},2R_{2}^{0}}\}$.\\
Finally, the estimates on the derivatives of $\widetilde{a}_{j}$ are elementary consequences of \eqref{Estimate for aj}. Indeed, 
\begin{align*}
\widetilde{\partial_{x_{1}} a_{j}}(z,w)=&\left(\partial_{z} \widetilde{a}_{j}+\partial_{w} \widetilde{a}_{j}\right)(z,w)\,,\qquad \widetilde{\partial_{x_{2}} a_{j}}(z,w)=i\left(\partial_{z} \widetilde{a}_{j}-\partial_{w} \widetilde{a}_{j}\right)(z,w)\,,\\
\widetilde{\Delta a_{j}}(z,w) =&4\partial_{z}\partial_{w} \widetilde{a}_{j}(z,w)\,,
\end{align*}
so that Cauchy estimates give, for all $(z,w)\in P(0;R_{1}^{0},R_{2}^{0})$,
\begin{align*}
&\left\vert \widetilde{\partial_{x_{1}} a_{j}}(z,w)\right\vert\leq \frac{R_{1}^{0}R_{2}^{0}(R_{1}^{0}+R_{2}^{0}) \Vert \widetilde{a}_{j} \Vert_{R_{1}^{0},R_{2}^{0}} }{\left(R_{1}^{0}-|z| \right)^2\left(R_{2}^{0}-|w| \right)^2}\,,\qquad \left\vert \widetilde{\partial_{x_{2}} a_{j}}(z,w)\right\vert\leq \frac{R_{1}^{0}R_{2}^{0}(R_{1}^{0}+R_{2}^{0}) \Vert \widetilde{a}_{j} \Vert_{R_{1}^{0},R_{2}^{0}} }{\left(R_{1}^{0}-|z| \right)^2\left(R_{2}^{0}-|w| \right)^2}\,,\\
&\left\vert \widetilde{\Delta a_{j}}(z,w)\right\vert\leq \frac{4R_{1}^{0}R_{2}^{0}\Vert \widetilde{a}_{j} \Vert_{R_{1}^{0},R_{2}^{0}} }{\left(R_{1}^{0}-|z| \right)^2\left(R_{2}^{0}-|w| \right)^2}\,.
\end{align*}
Then, by using theses estimates \eqref{Estimate for aj} for all $(z,w)\in P\left(0;\frac{R_{1}^{0}}{2},\frac{R_{2}^{0}}{2}\right)$, this concludes the proof.
\end{proof} 
 
\subsubsection{Proof of Theorem \ref{Theorem exponential}}
Let us recall \eqref{Phase S}. Since $Q$ is positive, there exist $\delta>0$ and $M_{1}, M_{2}>0$ such that
\begin{equation}\label{Quadratic}
M_{1} |x|^2\leq \Re P(x) \leq M_{2} |x|^2,\qquad \text{for all }x\in D(0,\delta)\,.
\end{equation}
By considering $\delta$ sufficiently small, we can assume that \eqref{Transport Est} holds. 

Let $\chi \in C_{c}^{\infty}(\R^2)$ be a smooth cut-off function which is equal to $1$ near $0$ and has support in a compact subset of $D(0,\delta)$.  We define our pseudomode as
\[ u_{h}(x)\coloneqq\chi(x) e^{-P(x)/h}\left[a_{0}(x)+\sum_{j=1}^{N(h)} h^{j}a_j(x)\right]\,,\qquad N(h)\coloneqq\lfloor (e m h)^{-1/7} \rfloor\,.\]
Here, $m$ is a constant appearing in \eqref{Transport Est} and $\lfloor \cdot \rfloor$ denotes the floor function. 

By construction in Theorem \ref{Theorem pseudomode}, for all $j\geq 1$, $a_{j}(0)=0$. Thus, for all $x\in D(0,\delta)$,
\begin{align*}
\left\vert a_{j}(x)\right\vert = \left\vert \int_{0}^{1} \frac{\dd}{\dd t} a_{j}(tx)\, \dd t\right\vert= \left\vert \int_{0}^{1} x\cdot \nabla a_{j}(tx)\, \dd t\right\vert \leq m^{j+1}j^{7j}|x|\,.
\end{align*}
Then, by using \eqref{Transport Est} and the definition of $N(h)$, we have, for some $C_1>0$,
\begin{equation}\label{eq.boundsumaj}
	\sum_{j=1}^{N(h)} h^{j} \vert a_{j}(x) \vert\leq   \sum_{j=1}^{N(h)}h^{j} m^{j+1} j^{7j} |x|\leq  \sum_{j=1}^{N(h)} m e^{-j}|x|\leq C_1|x|\,.
\end{equation}
Since $a_{0}(0)=1$, there exists $C_{2}>0$ (independent of $h$) such that, in a neighbourhood of $0$,
\[\left\vert a_{0}(x)+\sum_{j=1}^{N(h)} h^{j}a_j(x)\right\vert \geq \vert a_{0}(x) \vert - C_{1} |x|\geq C_{2}\,. \]
By using \eqref{Quadratic}, we get
\begin{align*}
\int_{\R^2} |u_{h}(x)|^2\, \dd x \gtrsim  h^{\frac12}\,.
\end{align*}
Let us now prove that for each $\varepsilon\in (0,1)$, $e^{\varepsilon P/h}\left( \mathscr{L}_{h,\A} - h\mu\right)u_{h}(x)$ is exponentially small. We write
\begin{align*}
&e^{\varepsilon P/h}\left( \mathscr{L}_{h,\A} - h\mu\right)u_{h}(x)\\
=&\underbrace{e^{\varepsilon P/h}\left[ \mathscr{L}_{h,\A},\chi\right]\left( e^{-P/h}\sum_{j=0}^{N} h^{j}a_j\right)}_{I_{h}}+\underbrace{\chi e^{\varepsilon P/h}\left( \mathscr{L}_{h,\A} - h\lambda\right)\left(e^{-P/h}\sum_{j=0}^{N} h^{j}a_j\right)}_{J_{h}}\,,
\end{align*}
where
\begin{align*}
\left[ \mathscr{L}_{h,\A},\chi\right] =-2h^2 \,\nabla \chi \cdot \nabla-h^2 \,\Delta \chi + 2h\,i\A \cdot \nabla \chi\,.
\end{align*}
The term $I_{h}$ is a function supported on $\mathrm{supp}(\nabla\chi)$:
\begin{equation*}
I_{h}
=e^{-\frac{(1-\varepsilon)P}{h}}\left[-2h^2 \nabla \chi\cdot \sum_{j=0}^{N(h)} h^{j}\nabla a_j +\left(-h^2 \Delta \chi+ 2h \nabla P \cdot \nabla \chi +2h\,i \A\cdot \nabla \chi\right)\left(\sum_{j=0}^{N(h)} h^{j}a_j\right) \right].
\end{equation*}
Thanks to \eqref{eq.boundsumaj}, we notice that $\sum_{j=0}^{N(h)}h^j a_j(x)$ is bounded uniformly in $x\in D(0,\delta)$ and in $h$. Similarly, $\displaystyle \sum_{j=0}^{N(h)} h^{j} \nabla a_{j}(x)$ is also uniformly bounded. 

Then, there exist $C,\tilde C>0$ such that, for all $h\in(0,h_{0})$ and for all $x\in D(0,\delta)$,
\[ \vert I_{h}(x)\vert\leq C e^{-(1-\varepsilon)M_{1} |x|^2/h} \textup{\textbf{1}}_{\textup{supp }\nabla \chi}(x)\leq C e^{-(1-\varepsilon)\tilde C/h}\,.\]
For $J_{h}$, we can notice that, by construction and by using Lemma \ref{lem.boundaj},
\begin{equation*}
\left\vert J_{h}(x) \right\vert = \left\vert \chi(x) e^{-(1-\varepsilon)P(x)/h}h^{N(h)+2} \Delta a_{N(h)} \right\vert  \lesssim  \left(h m N(h)^{7} \right)^{N(h)} \lesssim  e^{-N(h)}\lesssim  e^{-C/h^{1/7}}\,,
\end{equation*}
where, in the last estimate, we used the fact that $N(h)>\frac{1}{(emh)^{1/7}}-1$. 
 
This concludes the proof of Theorem~\ref{Theorem exponential}.

\subsection{Examples}\label{Sec.examples}
Of course, Theorem \ref{Theorem exponential} is only interesting if one can ensure that $\Gamma$ is not empty. Let us discuss conditions on 
$\mathbf{A}$ and $\mathbf{B}$ to ensure that $0\in\Gamma$.

First, since $\partial_{\overline{z}} \B(0)\neq 0$, we observe that $0$ cannot be a critical point of $\mathbf{B}$ at $0$. Then, also since $\Im \A (0)=0$, we write
\begin{align*}
	\B(x_{1},x_{2})
	&=a+bx_{1} + cx_{2}+\mathcal{O}\left(\vert x \vert^2\right)\,,\\
	\Im A_{1}(x_{1},x_{2}) &= d x_{1} + e x_{2} +\mathcal{O}\left(\vert x \vert^2\right)\,,\\
	\Im A_{2}(x_{1},x_{2}) &= f x_{1} + g x_{2} +\mathcal{O}\left(\vert x \vert^2\right)\,,
\end{align*}
in a neighbourhood of $0$, where $a,b,c\in \C$ and $d,e,f,g\in \R$. Below, we will denote $a_{1},b_{1}, c_{1}$ (respectively,  $a_{2},b_{2}, c_{2}$) as the real parts (respectively, imaginary parts) of $a,b,c$. Let us find conditions on these coefficients so that $0\in \Gamma$.
We have
\begin{align*}
	&\B(0)=a_{1}+ia_{2}\,,
	\hspace{0.5 cm} \partial_{z}\B(0)=\frac{1}{2}\left[b_{1}+c_{2}+i(b_{2}-c_{1})\right]\,, \hspace{0.5 cm}\partial_{\overline{z}}\B(0)=\frac{1}{2}\left[b_{1}-c_{2}+i(b_{2}+c_{1})\right]\,,\\
	&\partial_{x_{1}} \Im A_{1}(0) = d\,, \hspace{0.5 cm} \partial_{x_{2}} \Im A_{1}(0)=e\,, \hspace{0.5 cm} \partial_{x_{1}} \Im A_{2}(0) = f\,, \hspace{0.5 cm}\partial_{x_{2}} \Im A_{2}(0)=g\,.
\end{align*}
Then, we obtain
\begin{align*}
 \frac{\partial_{z} \B}{\partial_{\overline{z}} \B}(0) =& \frac{b_{1}^2+b_{2}^2-c_{1}^2-c_{2}^2-i2(b_{1}c_{1}+b_{2}c_{2})}{(b_{1}-c_{2})^2+(b_{2}+c_{1})^2}\,,\\
\Re\left[ B(0) \frac{\partial_{z} \B}{\partial_{\overline{z}} \B}(0)\right] =& \frac{a_{1}\left(b_{1}^2+b_{2}^2-c_{1}^2-c_{2}^2\right)+2a_{2}(b_{1}c_{1}+b_{2}c_{2})}{(b_{1}-c_{2})^2+(b_{2}+c_{1})^2}\,,\\
\Im\left[ B(0) \frac{\partial_{z} \B}{\partial_{\overline{z}} \B}(0)\right] =& \frac{a_{2}\left(b_{1}^2+b_{2}^2-c_{1}^2-c_{2}^2\right)-2a_{1}(b_{1}c_{1}+b_{2}c_{2})}{(b_{1}-c_{2})^2+(b_{2}+c_{1})^2}\,,
\end{align*}
and thus,
\begin{align*}
Q_{1}(0) =& \frac{1}{2}\left[\frac{a_{1}\left(b_{1}^2+b_{2}^2-b_{1}c_{2}+b_{2}c_{1}\right)+a_{2}(b_{1}c_{1}+b_{2}c_{2})}{(b_{1}-c_{2})^2+(b_{2}+c_{1})^2} +d\right]\,,\\
Q_{2}(0) =& \frac{1}{2}\left[\frac{a_{2}\left(b_{1}^2+b_{2}^2-b_{1}c_{2}+b_{2}c_{1}\right)-a_{1}(b_{1}c_{1}+b_{2}c_{2})}{(b_{1}-c_{2})^2+(b_{2}+c_{1})^2} +e\right]\,,\\
Q_{3}(0) =& \frac{1}{2}\left[\frac{a_{1}\left(c_{1}^2+c_{2}^2-b_{1}c_{2}+b_{2}c_{1}\right)-a_{2}(b_{1}c_{1}+b_{2}c_{2})}{(b_{1}-c_{2})^2+(b_{2}+c_{1})^2} +g\right]\,.
\end{align*}
Here, in $Q_{2}(0)$, we have used $f-e=a_{2}$ deduced from $\partial_{x_{1}} \Im A_{2}(0) - \partial_{x_{2}} \Im A_{1}(0) = \Im \B(0)$.\\
Then, $0\in \Gamma$ if and only if
\begin{align}
 &a_{1}\neq 0 \text{ or } a_{2} \neq 0\,, \label{Assump B not 0}\\
 &b_{1}\neq c_{2} \text{ or }b_{2}\neq -c_{1}\,,\label{Assump dB not 0}\\
 &Q_{1}(0)>0\,,\label{Assump Q1>0}\\
 &Q_{1}(0)Q_{3}(0)-Q_{2}^2(0)>0\,. \label{Assump Q123>0}
\end{align}
\begin{example}[Polynomial complex magnetic fields]
In this example, we want to provide a class of polynomial $\B$ such that $\mathscr{L}_{h,\A}$ is well defined and $0\in \Gamma$. We consider
	\[ \B(x_{1},x_{2})=a+bx_{1}+cx_{2}+R(x_{1},x_{2})\,,\]
where
\begin{enumerate}[$\bullet$]
\item $a,b,c\in \C$ such that $a_{1}>0$, $a_{2}=0$ and $b_{2}c_{1}-b_{1}c_{2}>0$\,,
\item $R:\R^2\to \R$ is a polynomial such that 
\[R(0)=\partial_{x_{1}} R(0)=\partial_{x_{2}}R(0)=0 \qquad\text{and} \qquad \vert x \vert^4=o \left(R(x)\right) \text{ as } \vert x \vert \to +\infty\,.\]
\end{enumerate}
Let us choose the following magnetic potential
\[\begin{aligned}
		A_{1}(x_{1},x_{2})= 0\,,\qquad		A_{2}(x_{1},x_{2})=\int_{0}^{x_{1}} \B(s,x_{2})\, \dd s\,.
	\end{aligned}\]
Since $R$ is real-valued, we have
\[ \Re \B(x_{1},x_{2}) =a_{1} + b_{1}x_{1}+c_{1}x_{2}+R(x_{1},x_{2})\,,\qquad \Im A_{2}(x_{1},x_{2}) = \frac{b_{2}}{2}x_{1}^2 + c_{2}x_{1}x_{2}\,.\]
From the asymptotic behaviour of $R$, we deduce that
	\[|\Im \A(x)|^2=o\left( \Re \B(x)\right)\,,\qquad \text{ as } \vert x \vert \to +\infty\,.\]
	Then, \eqref{bound by Re B} is satisfied (and thus Assumption \ref{assumption.1} holds). This allows us to define $\mathscr{L}_{h,\mathbf{A}}$ by Theorem \ref{Thrm Operator}. Furthermore, since $|\Re \mathbf{B}(x)|\to +\infty$ as $|x|\to+\infty$, the operator has discrete eigenvalues by Theorem \ref{Thrm Copact Res}.

Since $a_{1}>0$ and $b_{2}c_{1}-b_{1}c_{2}>0$, \eqref{Assump B not 0} and \eqref{Assump dB not 0} hold.  Note that $d=e=g=0$ and $a_{2}=0$, we have
\begin{align*}
Q_{1}(0) =& \frac{1}{2}\frac{a_{1}\left(b_{1}^2+b_{2}^2+b_{2}c_{1}-b_{1}c_{2}\right)}{(b_{1}-c_{2})^2+(b_{2}+c_{1})^2}\,,\\
Q_{2}(0) =& -\frac{1}{2}\frac{a_{1}\left(b_{1}c_{1}+b_{2}c_{2}\right)}{(b_{1}-c_{2})^2+(b_{2}+c_{1})^2} \,,\\
Q_{3}(0) =& \frac{1}{2}\frac{a_{1}\left(c_{1}^2+c_{2}^2+b_{2}c_{1}-b_{1}c_{2}\right)}{(b_{1}-c_{2})^2+(b_{2}+c_{1})^2}\,.
\end{align*}
Obviously, $Q_{1}(0)>0$. Let us compute 
\begin{align*}
Q_{1}(0)Q_{3}(0)-Q_{2}(0)^2= \frac{a_{1}^2\left[\left(b_{1}^2+b_{2}^2+b_{2}c_{1}-b_{1}c_{2}\right)\left(c_{1}^2+c_{2}^2+b_{2}c_{1}-b_{1}c_{2}\right)-\left(b_{1}c_{1}+b_{2}c_{2}\right)^2 \right]}{4\left[(b_{1}-c_{2})^2+(b_{2}+c_{1})^2 \right]^2}\,.
\end{align*}
Notice that
\begin{align*}
&\left(b_{1}^2+b_{2}^2+b_{2}c_{1}-b_{1}c_{2}\right)\left(c_{1}^2+c_{2}^2+b_{2}c_{1}-b_{1}c_{2}\right)-\left(b_{1}c_{1}+b_{2}c_{2}\right)^2 \\
=& \left(b_{2}c_{1}-b_{1}c_{2}\right)\left[ (b_{1}-c_{2})^2+(b_{2}+c_{1})^2\right]\,.
\end{align*}
This yields that
\[ Q_{1}(0)Q_{3}(0)-Q_{2}(0)^2 =\frac{a_{1}^2\left(b_{2}c_{1}-b_{1}c_{2}\right)}{4(b_{1}-c_{2})^2+(b_{2}+c_{1})^2 }>0\,.\]
Thus, we have $0\in\Gamma$. By Theorem \ref{Theorem exponential}, there exist a pseudomode $u(\cdot,h)$ such that
\begin{equation}
		\left\Vert \left( \mathscr{L}_{h,\A} - h a_{1}\right)u_{h}(x)\right\Vert \leq  \exp\left(-\frac{C}{h^{1/7}}\right) \Vert u_{h}(x) \Vert\,,\qquad \text{as }h\to 0\,.
	\end{equation}
\end{example}
\begin{example}[Bounded oscillating magnetic fields]
	Consider  the magnetic potential
	\[ A_{1}(x_{1},x_{2})=-\sin(x_{1})x_{2}+i\cos(x_{2})\,,\qquad A_{2}(x_{1},x_{2})=i\cos(x_{2})\,.\]
	We have
	\[ \B(x_{1},x_{2})=\sin(x_{1})+i \sin(x_{2})\,.\]
	Since $\Im \A$ is bounded, both conditions \eqref{bound by Re B} and \eqref{bound by Im B} hold, therefore, the magnetic Laplacian $\mathscr{L}_{h,\A}$ is well-defined by Theorem \ref{Thrm Operator}. 
	Let us find the set $\Gamma$ in this explicit example. Take $x^{0}=(x_{1},x_{2})\in \Gamma$, we have 
\begin{align*}
\Im \A(x^{0})=0 \Longleftrightarrow \cos(x_{2})=0\,.
\end{align*}	
Then, it implies that $\B(x^{0})\neq 0$ and
\[\partial_{\overline{z}}\B(x^{0}) = \partial_{z} \B(x^{0})= \frac{1}{2}\cos(x_{1})\, .\]
Therefore, it is obvious that
\[\partial_{\overline{z}}\B(x^{0}) \neq 0 \Longleftrightarrow \cos(x_{1})\neq 0\,. \]
By straightforward calculation, we obtain
\[ Q_{1}(x^{0})=\frac{1}{2}\sin(x_{1}),\qquad Q_{2}(x^{0})=0,\qquad Q_{3}(x^{0})=-\frac{1}{2}\sin(x_{2})\,.\]
Therefore, it yields that
\begin{align*}
Q_{1}(x^{0})>0 &\Longleftrightarrow \sin(x_{1})>0\,,\\
Q_{1}(x^{0})Q_{3}(x^{0})-Q_{2}^2(x^{0})>0 &\Longleftrightarrow \sin(x_{2})=-1\,.
\end{align*}
	 The above analysis leads to
	 \begin{align*}
	 \Gamma &= \{ (x_{1},x_{2})\in \R^2:\sin(x_{1})>0,\hspace{0.2 cm} \cos(x_{1})\neq 0, \hspace{0.2 cm} \cos(x_{2})=0,\hspace{0.2 cm} \sin(x_{2})=-1\}\\
	 &= \left\{\left(0,\pi \right)\setminus \left\{\frac{\pi}{2} \right\}+2k\pi: k\in \Z\right\}\times\left\{ -\frac{\pi}{2}+ 2\pi n: n\in \Z \right\}\,.
 \end{align*}
 Then, Theorem \ref{Theorem exponential} states that: for each $x_{1}^{0} \in \left(0,\pi \right)\setminus \left\{\frac{\pi}{2} \right\}+2k\pi$ for some $k\in \Z$, we can construct a pseudomode $u(\cdot,h)$ such that
\begin{equation}
		\left\Vert \left( \mathscr{L}_{h,\A} - h (\sin(x_{1}^{0})-i)\right)u_{h}(x)\right\Vert \leq  \exp\left(-\frac{C}{h^{1/7}}\right) \Vert u_{h}(x) \Vert\,,\qquad \text{as }h\to 0\,.
	\end{equation}
\end{example}

\appendix

\section*{Acknowledgements}
T.N.D. would like to express his heartfelt gratitude to his late mother, Thi Quy Nguyen, whose love and support have been a constant source of inspiration. This work is dedicated to her memory. D.K. and T.N.D. were supported by the EXPRO grant number 20-17749X of the Czech Science Foundation (GA\v{C}R). N.R. and T.N.D. are grateful to CIRM where this work was started.

\bibliographystyle{amsplain}
\bibliography{Ref06}
\end{document}